\documentclass[12pt]{article}
\usepackage{epsfig}
\usepackage{graphicx}
\usepackage{amsmath, amsfonts, amsthm,amssymb}
\usepackage{latexsym}
\usepackage{color}
\usepackage{mathrsfs}
\usepackage{natbib}
\usepackage{hyperref}
\usepackage{appendix}
\usepackage{dsfont}
\usepackage[left=3cm,top=4cm,right=3cm,bottom=3cm]{geometry}
\RequirePackage[caption=false]{subfig}
\usepackage{lscape}
\usepackage{dsfont}
\usepackage{rotating}

\AtBeginDocument{%
}

\newcommand{\eqaldot}{\mathrel{=\mkern-0.25mu:}}
\usepackage{mathabx}
\usepackage{calligra}

\graphicspath{{./figs/}}

\newcommand{\de}{\mathrm{d}}
\newcommand{\e}{\mathds{E}}
\newcommand{\var}{\mathds{V}\text{ar}}
\newcommand{\cov}{\mathds{C}\text{ov}}
\newcommand{\corr}{\mathds{C}\text{or}}

\newcommand{\knl}{\mathfrak{K}}

\DeclareMathOperator{\card}{card}

\DeclareMathOperator{\tf}{\mathfrak{t}}

\newcommand{\zo}{\mathbf{z}(\circ)}

\begin{document}
\title{\LARGE \bf Summary characteristics for multivariate  function-valued spatial point process attributes}
\date{\date{}}
\maketitle
\begin{center}
{{\bf Matthias Eckardt$^{a}$, Carles Comas$^{b}$} and {\bf Jorge Mateu}$^{c}$}\\
\noindent $^{\text{a}}$ Chair of Statistics, Humboldt-Universit\"{a}t zu Berlin, Spandauer Strasse 1 , Berlin, Germany\\
\noindent $^{\text{b}}$ Department of Mathematics, Universitat de Lleida, Av. Alcalde Rovira Roure 191, 
 Lleida, Spain\\
\noindent $^{\text{c}}$ Department of Mathematics, Universitat  Jaume I, E-12071, Castell{\'o}n, Spain\\
\end{center}
\begin{abstract}
Prompted by modern technologies in data acquisition, 
the statistical analysis of spatially distributed function-valued quantities has attracted a lot of attention in recent years. In  particular, combinations of functional variables and spatial point processes yield a highly challenging instance of such modern spatial data applications. Indeed, the analysis of spatial random point configurations, where the point attributes themselves are functions rather than scalar-valued quantities, is just in its infancy, and  extensions to function-valued quantities still remain limited. In this view, we extend current existing first- and second-order summary characteristics for real-valued point attributes to the case where in addition to every spatial point location a set of distinct function-valued quantities are available. Providing
a flexible treatment of more complex point process scenarios, we build a framework to consider points with multivariate function-valued marks, and develop sets of different cross-function (cross-type and also multi-function cross-type) versions of summary characteristics that allow for the analysis of highly demanding modern spatial point process scenarios. We consider estimators of the theoretical tools and analyse their behaviour through a simulation study and two real data applications.

\end{abstract}

\section{Introduction}

Introducing general ideas from functional data analysis \citep{ramsey2005, ferraty2006nonparametric, horvath2012, Hsing2015} into the field of spatial statistics, the statistical analysis of functional spatial data has attracted a lot of attention in recent years \citep[see][for a general review]{Delicado2008,Mateu2017, 10.1214/20-BJPS466}. Potential applications from the literature include the analysis of  regional penetration
resistance profiles \citep{Giraldo2011}, air pollution monitoring data \citep{Bohorquez2017} and regional gross domestic product dynamics \citep{fsar}. Different from more classical spatial statistics \citep[see e.g.][]{cressie93}, the objects of interest in any such data are the realised trajectories, i.e. curves, of some underlying continuous mechanism which are collected over some spatial domain $\mathcal{X}\subset\mathds{R}^d$, usually $d=2$. As such, the  functional observations themselves are assumed to be positively (resp. negatively) dependent over space relative to the distance between the spatial entities which needs to be accounted for in any statistical analysis. To this end, various approaches  from classical spatial statistics were extended to functional outcomes yielding an ever-increasing methodological toolbox of different  functional spatial data analysis techniques.  Predetermined by the exact nature of $\mathcal{X}$, these  techniques help to investigate the spatial interrelations between the individual functional objects in geostatistical, areal or point process data contexts. However, despite a relatively large body of contributions on the analysis of geostatistical functional data and corresponding functional kriging approaches \citep[see][for general reviews]{doi:https://doi.org/10.1002/9781119387916.ch13, doi:https://doi.org/10.1002/9781119387916.ch3,  
doi:https://doi.org/10.1002/9781119387916.ch4}, where a functional outcome is collected over a set of fixed locations,  
and a growing number of contributions to functional areal data  \citep{fcar,fsar, AwCabral2020}, the analysis of spatial random  point configurations, where the point attributes themselves are functions rather than scalar-valued quantities, is just in its infancy. 

Despite the notable progress in spatial point process methodology with extensions to more challenging non-Euclidean domains for the points including the sphere, linear networks and graphs with Euclidean edges, extensions to more complicated non-scalar marks have not been covered much so far. 
While the mark covariance \citep{DBLP:journals/eik/Stoyan84}, mark correlation \citep{isham1985marked,StoyanStoyan1994}, mark weighted $K$ \citep{pettinen1992forest}, mark variogram \citep{cressie93,markvar,Stoyan2000}, and mark differentiation \citep{20113358596,HUI2014125} functions, or corresponding nearest-neighbour versions \citep{StoyanStoyan1994}, have become prominent tools for the  characterisation of real-valued marks,  extensions to function-valued quantities still remain limited. In contrast to (marked) spatio-temporal point processes  \citep{10025364994, VereJones2009,GONZALEZ2016505}  where different clustered point process \citep{GONZALEZ2016505}, 
Gibbsian processes \citep{RENSHAW200185,SARKKA20061698, RENSHAW2009, REDENBACH2013672}, log-Gaussian \citep{Serra2014, Siino2018} and shot-noise \citep{brix2002, Moller2010FIRE} Cox model specifications, or corresponding second-order summary characteristics \citep{https://doi.org/10.1111/j.1467-9574.2008.00407.x, STOYAN2017125, https://doi.org/10.1111/sjos.12367} used to characterise the temporal evolution for a set of (marked) point locations, contributions to function-valued marks remain elusive. In particular, advances to sets of distinct function-valued attributes, i.e. multivariate curves, do not exist.

Originating in the pioneering paper of \cite{comas2008METMA} and subsequent works by \citet{Comas2011, Comas2013} which introduced and  extended the mark correlation function for function-valued point attributes, \citet{Ghorbani2020} were the first to provide a mathematically rigorous treatment on the subject. Instead of the set $\lbrace(\mathbf{x}_i, m(\mathbf{x}_i))\rbrace^n_{i=1}$
with points $\mathbf{x}_i\in \mathcal{X}$ and  scalar-valued marks $m(\mathbf{x}_i)$ on some suitable mark space $\mathds{M}$, these authors considered the set   $\lbrace(\mathbf{x}_i, (f(\mathbf{x}_i), l(\mathbf{x}_i)))\rbrace^n_{i=1}$ where each point $\mathbf{x}_i$ is augmented by a function-valued quantity $f(\mathbf{x}_i)\in \mathds{F}$ and, potentially, an additional Euclidean auxiliary mark $l(\mathbf{x}_i)$ living on a suitable latent mark space $\mathds{L}$. As such, apart from the trivial case when no auxiliary mark is available, this formulation allows for (i) function-valued marked  multivariate point patterns, where different types of points with one function-valued point attribute are observed, and (ii) function-valued marked point patterns where at each point additional real-valued information is available. While  providing a flexible treatment of more complex point process scenarios which include unmarked (resp. scalar-valued marked) point processes as special case when the $(f(\mathbf{x}_i),l(\mathbf{x}_i))$ (resp. $f(\mathbf{x}_i)$) argument is ignored, extensions to points with multiple distinct function-valued marks have not been covered to the very best of our knowledge. Considering at least two distinct function-valued point attributes for each point location, this paper aims to fill this gap. In particular, sets of different cross-function summary characteristics for points with two distinct function-valued marks are introduced and extended to cross-type and also multi-function cross-type versions that allow for the analysis of highly demanding modern spatial point process scenarios.  All data and R code to reproduce the proposed auto- and cross-function mark characteristics is made publicly available in a github repository \url{https://github.com/carlescomas/SppFDA}.      


The remainder of the paper is structured as follows. After a general introduction to spatial point processes with multivariate function-valued point attributes, Section \ref{sec:methods} establishes different cross-function mark characteristics and potential extension to multitype point processes. In particular, extensions of classical test functions  to the function-valued mark setting are discussed in  Section \ref{sec:testfunction}. Estimators of the proposed mark characteristics are presented in Section \ref{sec:estimates}. The proposed characteristics are evaluated through a simulation study in Section \ref{sec:sim}. Section \ref{sec:appl} presents an application of the proposed tools to two different data sources originating from forestry and urban economic contexts. The paper concludes with a discussion in Section \ref{sec:final}.

\section{Spatial point processes with multivariate function-valued marks}\label{sec:methods}

To extend the theory and methodology of  function-valued marked spatial point processes to  multivariate function-valued point attributes, let $\mathcal{X}$ denote a subset of $\mathds{R}^2$ equipped with Borel sets $\mathcal{B}(\mathcal{X})$, and $d(\cdot)$ an Euclidean metric on $\mathcal{X}$. On $\mathcal{X}$, define $\Psi_G=\lbrace \mathbf{x}_i \rbrace^n_{i=1}$  as ground, i.e. unmarked,  spatial point process with intensity measure $\Lambda_G$. As such, $\Psi_G$ is well embedded into the theory of spatial point processes  and a rich body of different tools can directly be applied to investigate the structural properties of the points  \citep[see][for  general references to spatial point processes]{moller2004,Illian2008,Chiu2013}. Associated with $\Psi_G$, denote by  $\Psi=\lbrace \mathbf{x}_i, \mathbf{f}(\mathbf{x}_i) \rbrace^n_{i=1}$ a marked spatial point process on $\mathcal{X}\times \mathds{F}^p$ with locations  $ \mathbf{x}_i\in\mathcal{X}$ and $p$-variate associated function-valued marks $\mathbf{f}(\mathbf{x}_i)=(f_1(\mathbf{x}_i), \ldots, f_p(\mathbf{x}_i))$ on $\mathds{F}^p$ where each $f_h(\mathbf{x}_i): \mathcal{T}\subseteq \mathds{R} \mapsto \mathds{R}, h=1,\ldots,p$  with $\mathcal{T}=(a,b), -\infty\leq a\leq b\leq \infty$. In general, $\mathds{F}^p$ is assumed to be a Polish, i.e. complete separable metric, space  equipped with $\sigma$-algebra $\mathcal{F}^p=\bigotimes_{h=1}^p \mathcal{F}_h$ \citep{Daley2008}. For $\Psi$, the expected number of points $N_h(\cdot)$ in  $B\in\mathcal{B(X)}$ with function-valued attribute in $F_h\in \mathcal{F}_h$ corresponds to the intensity measure  $\Lambda_h(B\times F_h)$ which simplifies to 
\[
\e\left[N_h(B\times F_h)\right]=\Lambda_h(B\times F_h)=\int_{B\times F_h}\lambda_G(\mathbf{x})d\mathbf{x}P(dF_h)
\]
for fixed $F_h$ in $\mathcal{F}_h$ with $\lambda_G$ being the intensity function of $\Psi_G$ and $P(dF_h)$ a reference measure on $(\mathds{F}^p,\mathcal{F}^p)$. For stationary $\Psi$, i.e. if $\Psi=\Psi_x$ with $\Psi_x=\lbrace (\mathbf{x}_i+x, \mathbf{f}(\mathbf{x}_i))\rbrace^n_{i=1}$ for any translation $x$, and fixed $f_h$ the intensity measure $\Lambda_h(B\times F_h)$ equals $ \lambda_h\nu(B)$ with $\lambda_h$ denoting the intensity of $\Psi$ with respect to $F_h$ and $\nu(\cdot)$ the Lebesgue measure, i.e. the volume, of its argument. Similarly, $\Psi$ is called isotropic if $\Psi=\mathfrak{r}\Psi$ with $\mathfrak{r}\Psi=\lbrace (\mathfrak{r}\mathbf{x}_i, \mathbf{f}(\mathbf{x}_i))\rbrace^n_{i=1}$ for any rotation $\mathfrak{r}$.

To account for additional integer-valued marks, i.e. when  different types of points are available,  $\Psi$ can be generalised to a $m$-variate (i.e. multitype) spatial point process $\boldsymbol{\Psi}$ with $n=n_1+\ldots+n_m$ points and multivariate function-valued point attributes on  $\mathcal{X}^m\times \mathds{F}^p$ with corresponding component processes  $\Psi_d,~d=1,\ldots,m$ and associated ground process $\boldsymbol{\Psi}_G$. We note that the above point processes could also be extended by additional real-valued mark information, e.g. through additional auxiliary mark terms $l(\mathbf{x}_i)\in\mathds{R}$, which allows for the formulation of doubly-marked (multitype) point processes where each point is augmented by multivariate function-valued and one (resp. $w$ distinct) real-valued marks living on $\mathcal{X}^m\times \mathds{F}^p\times\mathds{R}$ (resp. $\mathcal{X}^m\times \mathds{F}^p\times\mathds{R}^w$).   
 
\subsection{Cross-function second-order mark summary characteristics and nearest neighbour indices}\label{sec:testfunction}

Apart from the first-order properties, a variety of second-order mark summary characteristics and their related nearest-neighbour versions have become useful methodological tools for the analysis of classical (real-valued) marked spatial point process scenarios which help to investigate the heterogeneity and interrelation between the observed point attributes, and decide on the independent mark hypothesis as a function of the distance between pairs of two points. To extend the methodological toolbox to the function-valued marks setting and define suitable cross-function characteristics, let $f_h(\mathbf{x})$ and $f_l(\mathbf{x}')$ denote two distinct function-valued marks for a pair of distinct point locations in $\Psi$ with interpoint distance  $d(\mathbf{x},\mathbf{x}')=r,~r>0$. Adopting the core principles from classical mark characteristics and applying a pointwise evaluation first, different cross-function mark characteristics can be defined by introducing a test function $\tf_f$ \citep{10.2307/4616131}, i.e. a map $\tf_f: \mathds{F}\times\mathds{F}\rightarrow \mathds{R}^+$, which itself takes the marks $f_h$ and $f_l$ for a pair of distinct points in $\Psi$ as arguments. In what follows, we assume $\Psi$ to be second-order stationary and isotropic such that the characteristics solely depend on the distance $r,~r>0$ and it suffices to consider the marks at the origin $\circ$ and the distance $\mathbf{r}$ where $d(\circ, \mathbf{r})=r$. Depending on the precise specification of $\tf_f$, different mark characteristics can be constructed by taking the expectation $\e_{\circ,\mathbf{r}}$ of $\tf_f$ under the  condition that $\Psi$ has indeed points at locations $\circ$ and $\mathbf{r}$. Generalising the most prominent test functions from the literature  to the function-valued marks setting, potential specifications of $\tf_f$ include 
$\tf_1=1/2(f_h(\circ)(t)- f_l(\mathbf{r})(t))^2$ and $\tf_2(f_h(\circ)(t), f_l(\mathbf{r})(t))=(\min(f_h(\circ)(t), f_l(\mathbf{r})(t)))/(\max(f_h(\circ)(t), f_l(\mathbf{r})(t)))$, which are based on the difference or the ratio between the pair of distinct marks, and 
$\tf_3=f_h(\circ)(t)\cdot f_l(\mathbf{r})(t), \tf_4=f_h(\circ)(t)$ and $\tf_5=f_l(\mathbf{r})(t)$, which are based on the product of the arguments \citep[see][for detailed discussion]{Schlather2001, Illian2008}.  Obviously, the above formulations include auto-mark characteristics as special cases for $h=l$. We note that apart from a concurrent setting, alternative pointwise test functions may be defined for the marks $f_h(t)$ and $f_l(s)$ with $s<t$.

\subsubsection{Cross-function variation and differentiation characteristics}

As  a first cross-function mark characteristic, the pointwise  cross-function mark variogram $\gamma_{hl}(r,t)$  which helps to investigate the strength and range of the   variation in the mark differences with respect to the distance $\mathbf{r}$ can be derived by taking the conditional expectation  $\e_{\circ,\mathbf{r}}$  of $\tf_1$. This pointwise characteristic can then be turned into a global  cross-function mark variogram $\gamma_{hl}(r)$ which corresponds to the $L_2$ metric by integrating $\gamma_{hl}(r,t)$ over $\mathcal{T}$, 
\[
\gamma_{hl}(r)=\int_{\mathcal{T}}\e_{\circ,r}\left[\tf_1(f_h(\circ)(t),f_l(\mathbf{r})(t))\right]\de t \eqaldot \int_{\mathcal{T}}\gamma_{hl}(r,t)\de t. 
\]
While the limit of this characteristic equals the non-spatial variance and the normalised version yields a straight line that is constantly one under the independent mark assumption, large values of this cross-function characteristic will indicate a strong heterogeneity between the function-valued attributes $f_h$ and $f_l$ at a distance $r$.

Different from the cross-function mark variogram, taking $\tf_3$ as argument of $\e_{\circ,\mathbf{r}}$ yields a pointwise cross-function version of Stoyan's mark covariance function \citep{DBLP:journals/eik/Stoyan84} defined by $\cov^{\mathrm{Sto}}_{hl}(r,t)=\e_{\circ,\mathbf{r}}\left[\tf_3\right]-\mu_h(t)\cdot\mu_l(t)$ with corresponding global characteristic  $\cov^{\mathrm{Sto}}_{hl}(r)=\int \cov^{\mathrm{Sto}}_{hl}(r,t)\de t$ where $\mu_h(t)=\e\left[f_h(t)\right]$ and $\mu_l(t)=\e\left[f_l(t)\right]$ are the non-spatial means of $f_h(t)$ and $f_l(t)$, respectively. Alternatively, a cross-function mark covariance can also be obtained  by rewriting Cressie's \citep{cressie93} covariance function $\cov^{\mathrm{Cre}}_{hl}$ as $\cov^{\mathrm{Cre}}_{hl}(r,t)=\e_{\circ,\mathbf{r}}\left[\tf_3\right]-\e_{\circ,\mathbf{r}}\left[\tf_4\right]\e_{\circ,\mathbf{r}}\left[\tf_5\right]$ where $\cov^{\mathrm{Cre}}_{hl}(r)=\int\cov^{\mathrm{Cre}}_{hl}(r,t)\de t$.  

Inserting the ratio of $\tf_2$ instead of the difference between the paired marks into the conditional expectation $\e_{\circ,\mathbf{r}}\left[\cdot\right]$ yields a pointwise cross-function version of the mark differentiation function \citep{20113358596,HUI2014125} $\tau_{hl}(r,t)$ for function-valued marks defined by  $\tau_{hl}(r,t)=1-\e_{\circ,\mathbf{r}}\left[\tf_2\right]$ 
with global characteristic $\tau_{hl}(r)=\int\tau_{hl}(r,t)\de t$. Obviously, values of $\tau_{hl}(r,t)$ equal or close to zero imply that the function-valued point attributes at the distance $r$ are equal or almost identical while increasing non-zero values indicate an  increase in heterogeneity of the marks.     

\subsubsection{Cross-function correlation characteristics}

Different from the difference and ratio based characteristics, an alternative set of cross-function mark characteristics can be defined through the product of the function-valued marks. Taking $\tf_3$ as argument of $\e_{\circ,r}$ yields a pointwise cross-function version of the conditional mean product of marks $c_{hl}(r,t)$ within a distance $r$ at $t\in\mathcal{T}$. We note that $c_{hl}(r,t)$ translates again into a global cross-function characteristic $c_{hl}(r)$ by integration of the pointwise one over $\mathcal{T}$,
\begin{equation}\label{eq:crossprodmarks}
c_{hl}(r)=\int_{\mathcal{T}} \e_{\circ,\mathbf{r}}\left[f_h(\circ)(t)\cdot f_l(\mathbf{r})(t)\right]\de t \eqaldot  \int_{\mathcal{T}} c_{hl}(r,t)\de t.    
\end{equation}
Further, normalising $c_{hl}(r,t)$ by $\mu_h(t)\cdot\mu_l(t)$, i.e. the product  of non-spatial means,  yields a cross-function pointwise version of Stoyan's mark cross-correlation function $\kappa_{hl}(r,t)$ \citep{markCross}  from which the global characteristic $\kappa_{hl}(r)$ follows analogous to \eqref{eq:crossprodmarks} by integration of $\kappa_{hl}(r,t)$ over $\mathcal{T}$. We note that Stoyan's mark covariance function is indeed  a linear transformation of the mark correlation function such that $\cov^{\mathrm{Sto}}_{hl}(r)$ and $\kappa_{hl}(r)$ are essentially the same  \citep{Schlather2001}.

Apart from Stoyan's mark correlation function, \cite{isham1985marked} and \cite{Beisbart:2000}   introduced two alternative mark correlation functions that could also be extended to the function-valued mark setting. 
\cite{Beisbart:2000}  proposed a simpler version of the above formulation of the mark correlation function  in which the product of the mark values is replaced by the normalised sum of marks. Using this formulation allows for a straightforward extension to a pointwise version for function-valued marks $f_h(\circ)(t)$ and $f_l(\mathbf{r})(t)$ defined by  
\[
\kappa^{\mathrm{Bei}}_{hl}(r,t)=\frac{f_h(\circ)(t)+f_l(\mathbf{r})(t)}{\mu_h(t)+\mu_l(t)}
\]
with $\kappa^{\mathrm{Bei}}_{hl}(r)=\int \kappa^{\mathrm{Bei}}_{hl}(r,t)\de t$. As the nominator approaches the product (resp. the sum) of means $\mu_h(t)$ and $\mu_l(t)$ as limits,  both $\kappa_{kl}$ and $\kappa_{kl}^{\mathrm{Bei}}$ are constantly equal to one for all $t\in \mathcal{T}$ in case of independent marks and positive or negative mark correlations could easily be identified by positive or negative deviations from one, respectively. Opposite to the above formulations, \cite{isham1985marked} introduced a different type of mark correlation function which is closely related to Pearson's correlation. Using the cross-function version of Cressie's mark covariance function, a pointwise cross-function analogue to Isham's mark correlation function can be defined as
\[
\corr^{\mathrm{Ish}}_{hl}(r,t)=\frac{\cov^{\mathrm{Cre}}_{hl}(r,t)}{\sqrt{\var_{hh}(r,t)}\cdot\sqrt{\var_{ll}(r,t)}}
\] 
where $\var_{hh}(r,t)=\e_{\circ,\mathbf{r}}\left[(f_h(\circ)(t)\cdot f_h(\mathbf{r})(t)\right]-\e_{\circ,\mathbf{r}}\left[f_h(\circ)(t)\right]\e_{\circ,\mathbf{r}}\left[f_h(\mathbf{r})(t)\right]$ and $\var_{ll}(r,t)$ defined analogous, and $\corr^{\mathrm{Ish}}_{hl}(r)=\int \corr^{\mathrm{Ish}}_{hl}(r,t)\de t$.

Apart from the extended cross-function mark correlation  characteristics outlined above, taking $\tf_4$ or $\tf_5$ as arguments of $\e_{\circ,\mathbf{r}}$ leads to pointwise \textbf{r}-mark functions $c_{h\bullet}(r,t)$ and $c_{\bullet l}(r,t)$, respectively, where $c_{h\bullet}(r,t)=c_{\bullet l}(r,t)$. As before, both pointwise \textbf{r}-mark functions translate into global characteristics $c_{h\bullet}(r)$ and $c_{\bullet l}(r)$ by integration of $c_{h\bullet}(r,t)$ and $c_{\bullet l}(r,t)$ over $\mathcal{T}$, respectively. Further, normalisation of  $c_{h\bullet}(r,t)$ and $c_{\bullet l}(r,t)$ by $\mu_h(t)$ and $\mu_l(t)$ yields the pointwise \textbf{r}-mark correlation functions $\kappa_{h\bullet}(r,t)$ and $\kappa_{\bullet l}(r,t)$, respectively, where  $\kappa_{h\bullet}(r)=\int\kappa_{h\bullet}(r,t)\de t$ and $\kappa_{\bullet l}(r)=\int \kappa_{\bullet l}(r,t)\de t$.

We note that the cross-function mark correlation and $\mathbf{r}$-correlation functions can also be used to define a counterpart version of the $U(r)$ function for function-valued marks, this $U(r)$ being the mean product of marks sited at distance $r$ apart,
\begin{equation}
U(r)=\int \lambda^2 g(r) \kappa_{hl}(r)\de a\de a', 
\end{equation}
where $\lambda\equiv \lambda_G$ is the intensity of the points, $g(r)$ the pair correlation function, and $a$ and $a'$ are two infinitesimal small areas containing points $\mathbf{x}$ and $\mathbf{x}'$ which are separated by a  distance $r$ \citep{Capobianco1998, Renshaw2002}. Including second-order summary characteristics for both the points and the function-valued marks, this characteristics accounts jointly for spatial variation of the point locations and the marks. Under the independent marks assumptions, $\kappa_{hl}(r)=1$ whereas $g(r)=1$ under the complete spatial randomness hypothesis, i.e. the homogeneous Poisson point process case. Alternative formulation of $U(r)$ can be achieved by substituting $\kappa_{hl}(r)$ by the $\mathbf{r}$-correlation functions $\kappa_{h\bullet}(r)$ and $\kappa_{\bullet l}(r)$, the mark variogram $\gamma_{hl}(r)$ and the mark differentiation function $\tau_{hl}(r)$, or alternatively by rewriting  $U(r)$ into polar coordinates allows for also for anisotopic behaviour. 

\subsubsection{Cross-function nearest-neighbour indices and $k$-nearest neighbour characteristics}

While the second-order cross-characteristics provide functional summary characteristics of the pairwise interrelations between the function-valued point attributes  against the distance $r$, nearest-neighbour indices are essentially numerical mark summary characteristics which help to quantify the local variation between the marks for a pair of nearest-neighbouring points. Similar to the previous sections, different cross-function nearest-neighbour characteristics can be constructed by taking the conditional expectation of particular test functions which, in turn, only consider the function-valued marks $f_h(\circ)(t)$ and $f_l(\zo)(t)$ at the origin $\circ$ and its nearest neighbouring point $\zo$ \citep{StoyanStoyan1994}. 

Rewriting the test function $\tf_3$ into a nearest-neighbour version  $\tf_3^{\mathrm{nn}}=f_h(\circ)(t)\cdot f_l(\zo)(t)$ and taking the conditional expectation $\e_{\circ,\zo}\left[\tf_3^{\mathrm{nn}}\right]$ leads to a pointwise cross-function nearest-neighbour mark product index $c_{hl}^\mathrm{nn}(t)$ from which a corresponding pointwise nearest-neighbour mark product correlation index $\kappa_{hl}^\mathrm{nn}(t)$ derives directly by  normalising $c_{hl}^\mathrm{nn}(t)$ by  the product of means $\mu_h(t)\cdot\mu_l(t)$. Likewise, taking the conditional expectation of $\tf_4^{\mathrm{nn}}= f_l(\zo)(t)$  yields a pointwise nearest-neighbour mark index $c_{\bullet,l}^\mathrm{nn}(t)$ which transforms into the pointwise nearest-neighbour mark correlation index by normalising $c_{\bullet,l}^\mathrm{nn}(t)$ by $\mu_l$. Similarly,  pointwise cross-function nearest neighbour mark variogram and mark differentiation indices $\gamma_{hl}^\mathrm{nn}(t)$ and $\tau_{hl}^\mathrm{nn}(t)$ can be constructed by taking the conditional expectation $\e_{\circ,\zo}$ of the test functions $\tf_1^{\mathrm{nn}}=1/2(f_h(\circ)(t)- f_l(\zo)(t))^2$ and $\tf_2^{\mathrm{nn}}=\min_{\mathrm{nn}} /\max_{\mathrm{nn}}$ where $\min_{\mathrm{nn}}= \min(f_h(\circ)(t),f_l(\zo)(t))$ and $\max_{\mathrm{nn}}=\max(f_h(\circ)(t),f_l(\zo)(t))$, respectively. We note that all pointwise indexes translate into global numerical summary characteristics by integration of the pointwise version over $\mathcal{T}$.

Apart from considering only the function-valued mark of the nearest neighbouring point location $\zo$, the nearest neighbour indices can also be used to compute cumulative cross-function $k$-nearest neighbour summary characteristics from the marks $f_h$ and $f_l$ at the origin $\circ$ and its $k$-th nearest neighbouring point $\mathbf{z}_v(\circ)$ with $v=1,\ldots,k$. Substituting $\zo$ by  $\mathbf{z}_v(\circ))$, a corresponding cumulative mark correlation index can be computed from the pointwise cross-function  $k$-th nearest neighbouring index $\mathcal{K}_k(t)$, 
\[
\mathcal{K}_k(t) = \left.\frac{1}{k}\mathds{E}_{\circ, \mathbf{z}_v}\left(\sum_{v=1}^k f_h(\circ)(t)\cdot f_l(\mathbf{z}_v(\circ))(t)\right)\right/\mu_h(t)\mu_l(t),
\]
as $\mathcal{K}_k=\int \mathcal{K}_k(t)\de t$.  Likewise, a cumulative mark variogram index can be defined as $\Gamma_k=\Gamma_k(t)\de t$ with 
\[
\Gamma_k(t)=\frac{1}{k}\mathds{E}_{\circ, \mathbf{z}_v}\left(\sum_{v=1}^k \frac{1}{2}(f_h(\circ)(t)-f_l(\mathbf{z}_v(\circ))(t))^2\right).
\]
In addition, a pointwise counterpart version of Hui's mark dominance index $D_k$ \citep{Hui1998} for function-valued marks can be defined as 
\[
D_k(t)=\frac{1}{k}\mathds{E}_{\circ, \mathbf{z}_v}\left(\sum_{v=1}^k\mathds{1}(f_h(\circ)(t)>f_l(\mathbf{z}_v(\circ))(t))\right)
\]
which translates into a global characteristics by computing $D_h=\int D_k(t) \de t$. 

\subsection{Cross-function mark-weighted summary characteristics}

A different useful set of cross-function cumulative summary characteristics can be defined by adjusting classical functional point process summary characteristics for the function-valued marks by introducing a test function as weight into the specific functional point process summary expression. Although the principal idea also applies for the empty space and nearest neighbour contact distribution functions and related quantities, we explicitly only cover extension of second-order summary characteristics to the function-valued mark scenario including the mark-weighted pair correlation, $K$ and $L$ functions.

\subsubsection{Mark-weighted characteristics for unitype  point processes with multivariate function-valued marks}

To define a suitable pair correlation function for function-valued marks $f_h$ and $f_l$ at locations $\mathbf{x}$ and $\mathbf{x}'$ in $\Psi$, let $\alpha_{\tf_f}^{(2)}(t)$ denote the pointwise cross-function second-order factorial moment measure with density $\varrho^{(2)}_{\tf_f}(t)$, i.e. the pointwise cross-function second-order product density functions. For $\tf_f=\tf_3$,  $\alpha_{hl}^{(2)}$  becomes  
\[
 \alpha^{(2)}_{hl}(B_1\times B_2)(t)=\sum^{\neq}_{\substack{(\mathbf{x}, f_h(\mathbf{x})),\\(\mathbf{x}', f_l(\mathbf{x}')) \in \Psi} }\e\left[\mathds{1}_{B_1}(\mathbf{x})\mathds{1}_{B_2}(\mathbf{x}')f_h(\mathbf{x})(t)\cdot f_l(\mathbf{x}')(t)\right]
\]
with density $\varrho^{(2)}_{hl}(t)$. Then, a pointwise mark-weighted pair correlation function $g_{hl}(r,t)$ can be defined as $g_{hl}(r,t)=\varrho^{(2)}_{hl}(r,t)/(\lambda^2\mu_h(t)\mu_l(t))$ and $g_{hl}(r)=\int g_{hl}(r,t)\de t$. Specifying $\tf_f=\tf_5$ instead, the pair correlation function equals $g_{\bullet l}(r,t)=\varrho^{(2)}_{\bullet l}(r,t)/(\lambda^2\mu_l(t))$ where $\varrho^{(2)}_{\bullet l}$ is the density of $\alpha^{(2)}_{\bullet l}(t)$.

Similarly, setting $\tf_f=\tf_3$ a cross-function pointwise extension of the mark-weighted $K$ function \citep{pettinen1992forest} can be  defined as 
\[
\lambda\mu_{h}(t)\mu_{l}(t) K_{hl}(r,t)=\e_{\circ}\left[\sum_{(\mathbf{x},f_l(\mathbf{x}))\in\Psi} f_h(\circ)(t)\cdot f_l(\mathbf{x})(t)\mathds{1}_{b(\circ,r)}\lbrace \mathbf{x}\rbrace \right],
\]
where $b(\circ,\mathbf{r})$ is a ball of radius $\mathbf{r}$ centred at the origin, $\lambda$ is the intensity of the points and $\mu_h(t)$ and $\mu_l(t)$ are the non-spatial means of $f_h(t)$ and $f_l(t)$, respectively. Likewise, for $\tf_4$ and $\tf_5$, the mark weighted pointwise $K$ function changes to   $\lambda\mu_{h}(t) K_{h\bullet}(r,t)=\e_{\circ}\left[\sum_{(\mathbf{x},f_l(\mathbf{x}))\in\Psi} f_h(\circ)(t)\mathds{1}_{b(\circ,r)}\lbrace \mathbf{x}\rbrace \right]$ and $\lambda\mu_{l}(t) K_{\bullet l}(r,t)=\e_{\circ}\left[\sum_{(\mathbf{x},f_l(\mathbf{x}))\in\Psi} f_l(\mathbf{x})(t)\mathds{1}_{b(\circ,r)}\lbrace \mathbf{x}\rbrace \right]$, respectively \citep[see][]{Illian2008}. 
Analogous to the classical scalar-valued case, the mark-weighted $L$ function $L_{\tf_f}(r,t)$ is preferable to use in practice instead of the mark-weighted $K_{\tf_f}(r,t)$ function, where 
$L_{\tf_f}(r,t)=\sqrt{K_{\tf_f}(r,t)/\pi}$. As for the pair correlation function, both $K_{\tf_f}(r,t)$ and $L_{\tf_f}(r,t)$ translate into global cross-function characteristics by integration of the pointwise versions over $\mathcal{T}$.  

Finally, applying the principal idea of the above mark-weighted second-order summary characteristics to local summary characteristics, a localised version of e.g. the pointwise cross-function mark-weighted $K$ function for the $u$-th point location of $\Psi$ can be defined for $\tf_f=\tf_3$ as
\[
\lambda\mu_h(t)\mu_l(t) K_u(r,t)= \e\left[\sum_{(\mathbf{x}',f_l(\mathbf{x}'))\in\Psi} f_h(\mathbf{x}_u)(t)\cdot f_l(\mathbf{x}'(t))  \mathds{1}_{b(\mathbf{x_u}, \mathbf{r})}\lbrace\mathbf{x}'\rbrace\right].
\]

\subsubsection{Extensions to multitype points with multivariate function-valued marks}\label{sec:multipoints}

In what follows, let $\Psi_i=\lbrace (\mathbf{x}_{i},\mathbf{f}(\mathbf{x}_i)\rbrace^{n_i}_{i=1}$ and $\Psi_j=\lbrace (\mathbf{x}_{j},\mathbf{f}(\mathbf{x}_j)\rbrace^{n_j}_{j=1}$ denote two component processes of $\boldsymbol{\Psi}$ where $i\neq j$ and $\mathbf{f}(\mathbf{x}_i)$ and $\mathbf{f}(\mathbf{x}_j)$ are $p$-variate function-valued marks on $\mathds{F}^p$.  For $\tf_f=\tf_3$ the pointwise cross-function second-order factorial moment measure can be rewritten into a cross-function cross-type measure $\alpha^{(2)}_{ij,hl}(t)$,
\[
 \alpha^{(2)}_{ij,hl}(B_1\times B_2)(t)=\sum^{\neq}_{\substack{(\mathbf{x}_i, f_h(\mathbf{x}_i))\in\Psi_i,\\(\mathbf{x}_j, f_l(\mathbf{x}_j)) \in \Psi_j}} \e\left[\mathds{1}_{B_1}(\mathbf{x}_i)\mathds{1}_{B_2}(\mathbf{x}_j)f_h(\mathbf{x}_i)(t)\cdot f_l(\mathbf{x}_j)(t)\right],
\] 
with density  $\varrho^{(2)}_{ij,hl}(B_1\times B_2)(t)$. This product density, in turn, allows to define a pointwise cross-function cross-type pair correlation function 
$g_{ij,hl}(r,t)=\varrho_{ij,hl}(t)/(\lambda_i\lambda_j\mu_h(t)\mu_l(t))$ with $g_{ij,hl}(r)=\int g_{ij,hl}(r,t)\de t$. Further, a
pointwise mark-weighted cross-function cross-type $K$ function can be defined as  
\[
\lambda_i\mu_h(t)\mu_l(t) K_{ij, hl}(r,t)=\e_{\circ,i}\left[\tf_f(f_h(\circ)(t), f_l(\mathbf{r}(t))\mathds{1}(b(\circ), r)\lbrace \mathbf{x}_j\rbrace \right]
\]
from which a pointwise cross-function dot-type version can be obtained through $\lambda K_{i\bullet, hl}=\sum_j \mu_h(t)\mu_l(t)\lambda_j K_{ij, hl}(r)$.





\subsection{Multiple point and function-valued marks}

Finally, while the previous sections were restricted to the derivation of mark summary characteristics defined through a set of different test functions with at most two distinct function-valued point attributes, we now discuss potential extensions to marked point process scenarios with $p\geq 3$ distinct function-valued marks. For simplicity, we restrict to the trivariate case which could be extended naturally to sets of $p$ with $p\geq 3$ distinct function-marks.

For points with function-valued marks $\mathbf{f}= (f_d, f_h, f_l)$, three different generalised test functions can be defined through unconditional and also conditional formulations. As a general first approach, function $f_d(t)$ could be related to the set $\lbrace f_h,f_l\rbrace$ by specifying $\tf(f_d(\circ)(t), \lbrace f_h,f_l\rbrace(\mathbf{r})(t))$. Using this formulation and writing $\mu_{hl}(\mathbf{r})(t)$ for the mean of functions $h$ and $l$ at $t\in\mathcal{T}$ and distance $\mathbf{r}$, trivariate versions of $\tf_1$ and $\tf_3$ can be expressed as   $\tf_1(f_d(\circ)(t), \lbrace f_h,f_l\rbrace(\mathbf{r})(t))=1/2(f_d(\circ)(t)-\mu_{hl})(\mathbf{r})(t))^2$  
and $\tf_3(f_d(\circ)(t), \lbrace f_h,f_l\rbrace(\mathbf{r})(t))=(f_d(\circ)(t)\cdot\mu_{hl})(\mathbf{r})(t))$. Instead of $\mu_{hl}$, alternative formulations might be defined through the sum of pairwise operations, e.g. 
\[
\tf_1(f_d(\circ)(t), \lbrace f_h,f_l\rbrace(\mathbf{r})(t))=1/2((f_d(\circ)(t)-f_{h})(\mathbf{r})(t))^2+(f_d(\circ)(t)-f_{h})(\mathbf{r})(t))^2).
\]

Instead of only the $h$-th and $l$-th function-valued marks, a second general approach could be defined by relating the $d$-th function to all three functions, i.e. $\tf(f_d(\circ)(t), \mathbf{f}(\mathbf{r})(t))$. Similar to the above formulation, suitable specifications might be defined through $\mu_{dhl}(\mathbf{r})(t)$, the mean of set $\mathbf{f}$ at $t\in\mathcal{T}$ and distance $\mathbf{r}$, or alternatively using a pairwise formulation which, in turn, would combine both auto- and cross-type terms.  
While both of the above versions are specified through unconditional formulations, an alternative approach might derive from the conditional function-valued marks $f_d|f_l$ and $f_h|f_l$ yielding $\tf(f_d|f_l(\circ)(t), f_h|f_l(\mathbf{r})(t))$. Both conditional marks could be derived by partialising out the effect of $f_l$ from $f_d$ and $f_h$ using e.g. standard functional regression methods. Although interesting, this conditional formulation will not be pursued in this paper to make it more concise and focused.

\section{Estimation of cross-function summary characteristics}\label{sec:estimates}

After having discussed extensions of various mark characteristics for (multitype) spatial point processes with multiple function-valued marks, their estimation from observed spatial point patterns is presented next. As before, the empirical cross-function estimators for unitype point patterns are first described. To this end, let $\psi$ denote a spatial point pattern of $n$ points observed in a bounded observation window $W$, where each point is augmented by a multivariate function-valued point attribute, and denote by $\boldsymbol{\psi}$ the corresponding multitype point pattern with multivariate function-valued marks and components $\psi_1,\ldots,\psi_d$. Further, let $\card(\cdot)$ denote the cardinality, i.e. the number of points, in the argument.  

\subsection{Estimation of cross-function summary characteristics}\label{sec:emp:crosssumm}

Using the results of Section \ref{sec:testfunction} and writing $c_{\tf}(r)$ to denote the conditional expectation for any specific test function $\tf_f$, both variation and product related cross-function mark summary characteristics can be derived through a generic function $\kappa_{\tf}(r)$ whose specific form itself depends on the specification of  $\tf_f$. Estimating the second-order $\tf_f$-product density $\varrho^{(2)}_{\tf}(r)$, and its analogue version  $\varrho^{(2)}(r)$ of the ground pattern $\psi_G$, by 
\begin{equation}\label{eq:rho:testfunction}
\widehat{\varrho}^{(2)}_{\tf}(r)=\frac{1}{2\pi r \nu(W)} \sum_{\substack{(\mathbf{x},f_h),\\(\mathbf{x}',f_l)~\in~\psi}}^{\ne}
\ell(\tf_f(f_h(\mathbf{x}),f_l(\mathbf{x}')))\knl_b(\Vert\mathbf{x}-\mathbf{x}'\Vert-r) e(\mathbf{x},\Vert \mathbf{x}-\mathbf{x}'\Vert ),
\end{equation}
and 
\begin{equation}\label{eq:rho:ground}
\widehat{\varrho}^{(2)}(r)=\frac{1}{2\pi r \nu(W)} \sum_{\mathbf{x},\mathbf{x}'\in \psi_G}^{\ne}
\knl_b(\Vert\mathbf{x}-\mathbf{x}'\Vert-r) e(\mathbf{x},\Vert\mathbf{x}-\mathbf{x}'\Vert),
\end{equation} 
respectively, where 
\[
\ell(\tf_f(f_h(\mathbf{x}),f_l(\mathbf{x}')))=\int_a^b\tf_f(f_h(\mathbf{x})(t),f_l(\mathbf{x}')
(t)) \de t
\]
and $\knl_b(\cdot)$ is kernel function with bandwidth $b$, $\nu(\cdot)$ the area of its argument, and $e(\cdot)$ is an edge correction factor,  then $\kappa_{\tf}(r)$ can be estimated by 
\begin{equation}\label{kappa_est}
\widehat{\kappa}_{\tf}(r)=\left.\frac{\widehat{\varrho}^{(2)}_{\tf}(r)}{\widehat{\varrho}^{(2)}(r)}\right/ \widehat{c}_{\tf},\quad\text{for~} r > 0,
\end{equation}
where $\widehat{c}_{\tf}=\sum_{\mathbf{x}}\sum_{\mathbf{x}'}\ell(\tf_f(f_h(\mathbf{x}),f_l(\mathbf{x}'))/n^2$ is an estimator of $c_{\tf}=c_{\tf}(\infty)$.

We note that $\kappa_{\tf}(r)$  can alternatively also be estimated by  
\begin{equation}\label{happa_est2}
\widehat{\kappa}_{\tf}(r)=\frac{1}{2\pi r \nu(W)} \sum_{\substack{(\mathbf{x},f_h),\\(\mathbf{x}',f_l)~\in~\psi}}^{\ne}\frac{\ell(\tf(f_h(\mathbf{x}),f_l(\mathbf{x}')))\knl_b(\Vert\mathbf{x}-\mathbf{x}'\Vert-r) e(\mathbf{x},\Vert\mathbf{x}-\mathbf{x}'\Vert)}{\widehat{\lambda}^2 \widehat{g}(r)\widehat{c}_{\tf}}
\end{equation}
where $\widehat{g}(r)=\widehat{\varrho}^{(2)}(r)/\widehat{\lambda}^2, r\ge 0,$ and $\widehat{\lambda}=\card(W)/\nu(W)$ are estimators for the pair correlation function and the intensity of the ground process, respectively.
Specifying $\tf_f$ by $\tf_1$ and $\tf_3$ in the above formulation of $\widehat{\kappa}_{\tf}$, the cross-function mark variogram and mark correlation function can be estimated by 
\begin{equation}\label{eq:estvario}
\widehat{\gamma}_{hl}(r)=\frac{1}{2\pi r \nu(W)} \sum_{\substack{(\mathbf{x},f_h),\\(\mathbf{x}',f_l)~\in~\psi}}^{\ne}\frac{\ell(\tf_1(f_h(\mathbf{x}),f_l(\mathbf{x})'))\knl_b(\|\mathbf{x}-\mathbf{x}'\|-r) e(\mathbf{x},\Vert \mathbf{x}-\mathbf{x}'\Vert )}{\widehat{\lambda}^2 \widehat{g}(r)\widehat{c_{\tf}}}.
\end{equation}
and 
\begin{equation}\label{eq:estmcor}
\widehat{\kappa}_{hl}(r)=\frac{1}{2\pi r \nu(W)} \sum_{\substack{(\mathbf{x},f_h),\\(\mathbf{x}',f_l)~\in~\psi}}^{\ne}\frac{\ell(\tf_3(f_h(\mathbf{x}),f_l(\mathbf{x})'))\knl_b(\|\mathbf{x}-\mathbf{x}'\|-r) e(\mathbf{x},\Vert \mathbf{x}-\mathbf{x}'\Vert )}{\widehat{\lambda}^2 \widehat{g}(r)\widehat{\mu_h}\widehat{\mu_l}},
\end{equation}
respectively, where $\widehat{\mu_l}$ is the empirical functional mean of mark $f_l$. Similarly, estimators for $\kappa_{h\bullet}(r)$ (resp. $\kappa_{\bullet l}(r)$) can be obtained by setting $\tf_f$ to $\tf_4$ (resp. $\tf_5$) and substituting $\widehat{\mu_h}$ (resp. $\widehat{\mu_l}$) for $\widehat{\mu_h}\widehat{\mu_l}$ in \eqref{eq:estmcor}.

\subsection{Estimation of cross-function nearest-neighbour indices}

Estimators for the cross-function nearest-neighbour indices can be derived analogous to Section \ref{sec:emp:crosssumm} by replacing the above test functions by the nearest neighbour counterpart versions. Using the nearest neighbour test functions  $\tf_1^{\mathrm{nn}}$ and $\tf_3^{\mathrm{nn}}$, the nearest-neighbour mark variogram and mark product correlation index can be estimated through
\[
\widehat{\gamma}^{\mathrm{nn}}_{hl}=\frac{1}{n}\sum^n_{i=1}\ell(\tf_1^{\mathrm{nn}}(f_h(\mathbf{x}_i),f_l(\mathbf{z}(i))))/\widehat{c_t}
\]
and
\[
\widehat{\kappa}^{\mathrm{nn}}_{hl}=\frac{1}{n}\sum^n_{i=1}\ell(\tf_3^{\mathrm{nn}}(f_h(\mathbf{x}_i),f_l(\mathbf{z}(i))))/\widehat{\mu_h}\widehat{\mu}_h,
\]
respectively. 

\subsection{Estimation of cross-function nearest-neighbour indices mark-weighted characteristics}

Estimators of the cross-function mark weighted $K$ function can be obtained by normalising the function $\widehat{k}_{hl}$, 
\[
\widehat{k}_{\tf}(r)=\sum^{\neq}_{\mathbf{x},\mathbf{x}'\in W} \frac{\ell(\tf_f(f_h(\mathbf{x})(t),f_l(\mathbf{x}')(t)))\mathds{1}\lbrace\Vert \mathbf{x}-\mathbf{x}'\Vert\leq r\rbrace}{\nu(W) }
\]
by the empirical versions of the intensity $\widehat{\lambda}^2$ and a suitable normalising factor $ \widehat{c_{\tf}}$ corresponding to the specific test function used with $\widehat{c_{\tf}}=\widehat{\mu}_h\widehat{\mu}_l$ for $\tf_f=\tf_3$. The normalised estimator can then be transformed in the corresponding cross-function mark weighted $L$ function by taking the square root of $\widehat{K}(r)$. Likewise, the cross-function mark weighted pair correlation function can be computed as $\widehat{g}_{\tf}(r)=\widehat{\varrho}^{(2)}_{\tf}(r)/\widehat{\lambda}^2\widehat{c_{\tf}}$ which becomes 
$\widehat{g}_{hl}(r)=\widehat{\varrho}^{(2)}_{hl}(r)/\widehat{\lambda}^2\widehat{\mu}_h\widehat{\mu}_l$ for choosing $\tf_3$ as test function.

For the multitype point pattern scenario with components $\psi_i$ and $\psi_j$ and function-valued marks $f_h$ and $f_l$, the cross-function cross-type mark weighted $K$ function can be estimated by dividing
\[
\widehat{k}_{ij,\tf}(r)=\sum^{\neq}_{\mathbf{x}_i,\mathbf{x}_j\in W} \frac{\ell(\tf_f(f_h(\mathbf{x}_i)(t),f_l(\mathbf{x}_j)(t)))\mathds{1}\lbrace\Vert \mathbf{x}_i-\mathbf{x}_j\Vert\leq r\rbrace}{\nu(W) }
\]
by $\widehat{\lambda}_i\lambda_j \widehat{c_{\tf}}$ with
$\widehat{\lambda}_i$ denoting the intensity of the $i$-th component of $\boldsymbol{\psi}$. Again, this function translates into the corresponding estimator of the $L$ function through the square root of $\widehat{K}$. Similarly to the unitype case, a cross-function cross-type mark weighted pair correlation function can be calculated by computing   $\widehat{g}_{ij,t}(r)=\widehat{\varrho}^{(2)}_{ij,\tf}(r)/\widehat{\lambda}_i\widehat{\lambda}_j\widehat{c}_{\tf}.$

\section{A simulation study}\label{sec:sim}

We conducted a simulation study to investigate  how our estimators of the cross-function summary characteristics behave not only under several point configurations (including random, cluster and regular structures) but, in particular, also under different mark scenarios including spatially independence and function-valued marks, and positive or negative  inter-dependencies between functions of type $h$ and type $l$. For the case when we have positive interaction between functions, functions of type $l$ grow or decrease when interacting with functions of type $h$, and vice versa, whilst for negative inter-dependencies, functions of type $l$ grow or decrease when interacting with functions of type $h$, but functions of type $h$ decrease or grow when interacting functions of type $l$. 

\subsection{Generating point patterns with function-valued  marks}

To control for the effect of the inherent point configuration  on the proposed estimators, we considered three distinct point process configurations including Poisson, cluster and regular point process scenarios. Each of these three cases were generated on the unit torus to avoid edge effects with an expected number of points $n=200$. To obtain a  clustered point process structure, we simulated a Thomas process \citep{Thomas} with offspring dispersion parameter $\sigma=0.04$, parent intensity $\lambda_p = 40$ and $\mu=5$ expected offsprings per parent yielding an average number of points of around $200$ in the unit square with a moderate clustered configuration.  The regular point process scenario was constructed using a Strauss process \citep{strauss, KellyRipley} with interaction effect parameter $q=0.05$, and a radius of interaction $R_{int}=0.025$ which ensures strong inhibition effects for short scales of interaction, with an average number of points of around $200$.

To generate spatial point patterns in which each point location is augmented by a set of function-valued quantities, we consider the continuous space–time stochastic process developed by \cite{RENSHAW200185}. 
In this model, marked points located on the unit torus grow
and interact with each other in terms of a suitable growth-interaction scheme. 
We adapt this approach to avoid point mortality and point immigration. In this way, 
we keep the same point pattern over time and their associated growth curves. Technically, the Renshaw and S{\"a}rkk{\"a} algorithm generates a spatial point pattern with function-valued marks $h$ and $l$ through

\begin{equation}\label{RS1}
\begin{array}{ll}
\begin{aligned}
f_{h}(\mathbf{x})(t+\de t)=
&f_{h}(\mathbf{x})(t)+\beta_h f_{h}(\mathbf{x})(t)(1-f_{h}(\mathbf{x})(t)/S_h)\de t\\
&+ \sum^{\neq}_{\substack{(\mathbf{x}, f_h(\mathbf{x})),\\(\mathbf{x}', f_l(\mathbf{x}')) \in \psi}}J_h(f_{h}(\mathbf{x})(t),f_{l}(\mathbf{x}')(t);\Vert \mathbf{x}-\mathbf{x}'\Vert) \de t
\end{aligned}
\end{array}
\end{equation}
and 
\begin{equation}\label{RS2}
\begin{array}{ll}
\begin{aligned}
f_{l}(\mathbf{x})(t+\de t)=
&f_{l}(\mathbf{x})(t)+\beta_l (1-f_{l}(\mathbf{x})(t)/S_l)\de t\\
&+ \sum^{\neq}_{\substack{(\mathbf{x}, f_l(\mathbf{x})),\\(\mathbf{x}', f_l(\mathbf{x}')) \in \psi}}J_l(f_{l}(\mathbf{x})(t),f_{h}(\mathbf{x}')(t);\Vert \mathbf{x}-\mathbf{x}'\Vert) \de t
\end{aligned}
\end{array}
\end{equation}
where $f_{h}(\mathbf{x})(t)$ and $f_{l}(\mathbf{x})(t)$ are two functions of point $\mathbf{x}$ at time $t$, $\psi$ is a realisation of $\Psi$, $\beta_h$ the intrinsic rate of growth, $S_h$ the non-spatial carrying capacity, $\|\mathbf{x}_1-\mathbf{x}_2\|$ the Euclidean distance between a pair of points, and
$J_h(\cdot)$ a suitable interaction function between points. Note that functions of type $h$ and $l$ grow in terms of the classic logistic growth and the immigration-death process, respectively. These two simple growth functions ensure that both functions remain bounded.

To generate positive correlation between functions, we use a Strauss like symmetric interaction function  \citep{RENSHAW200185} adapted to the case where the interaction is between the marks $f_{h}$ and $f_{l}$, 

\begin{equation}\label{Int-RS}
J(f_{h}(\mathbf{x})(t),f_{l}(\mathbf{x}')(t);\Vert \mathbf{x}-\mathbf{x}'\Vert)=
       \left\{
         \begin{array}{ll}
       c & \mbox{if $\|\mathbf{x}-\mathbf{x}'\|<D$}\\
       0 & \mbox{otherwise},
      \end{array}
    \right.
\end{equation}
where $c\in \mathbb{R}$ is a constant interaction effect. Here, points start to interact with each other with
constant value $c$ as soon as their distance is less than $D$. To ensure a symmetric interacting  structure, $J_h(\cdot)=J_l(\cdot)=J(\cdot)$, as  smaller function values affect the growth of larger ones 
 in the same way as larger function values affects smaller ones. To avoid interacting effects to decrease function values, we set $c>0$; $c<0$ implies function reduction and eventually negative function values. 
 
 Moreover, to generate negative correlation between functions,
 we take $J_h(\cdot)=J(\cdot)$ and $J_l(\cdot)=0$. Now functions of type $h$ take an advantage when interacting with functions of type $l$ (faster grow), whilst functions of type $l$ are not affected by the interaction with functions of type $h$. This promotes  an asymmetric function interaction, resulting in negative spatial correlation between growth functions of distinct type. 

To generate spatial point patterns with function-valued marks, we consider expressions (\ref{RS1}) and  (\ref{RS2}) with growth carrying capacity $S_h=S_l=5$, intrinsic rates of growth $\beta_h=0.05$, $\beta_l=0.2$ and interaction distance $D=0.05$. These scenario parameters are chosen as they give rise to functions that are convenient as illustrative examples. Moreover, to obtain the desired marked point patterns with spatially independent and/ or positive correlation between function-valued marks, we consider the interaction mechanism (\ref{Int-RS}) for  $J_h(\cdot)=J_l(\cdot)=J(\cdot)$, with interaction parameter $c=0$ and $c=0.5$, respectively. Whilst to generate negative spatial correlation between functions we assume the same interaction function (\ref{Int-RS}), but for  $J_h(\cdot)=J(\cdot)$ and $J_l(\cdot)=0$ with $c=0.5$.

\begin{figure}[!h]
    \centering
    \includegraphics[scale=0.3]{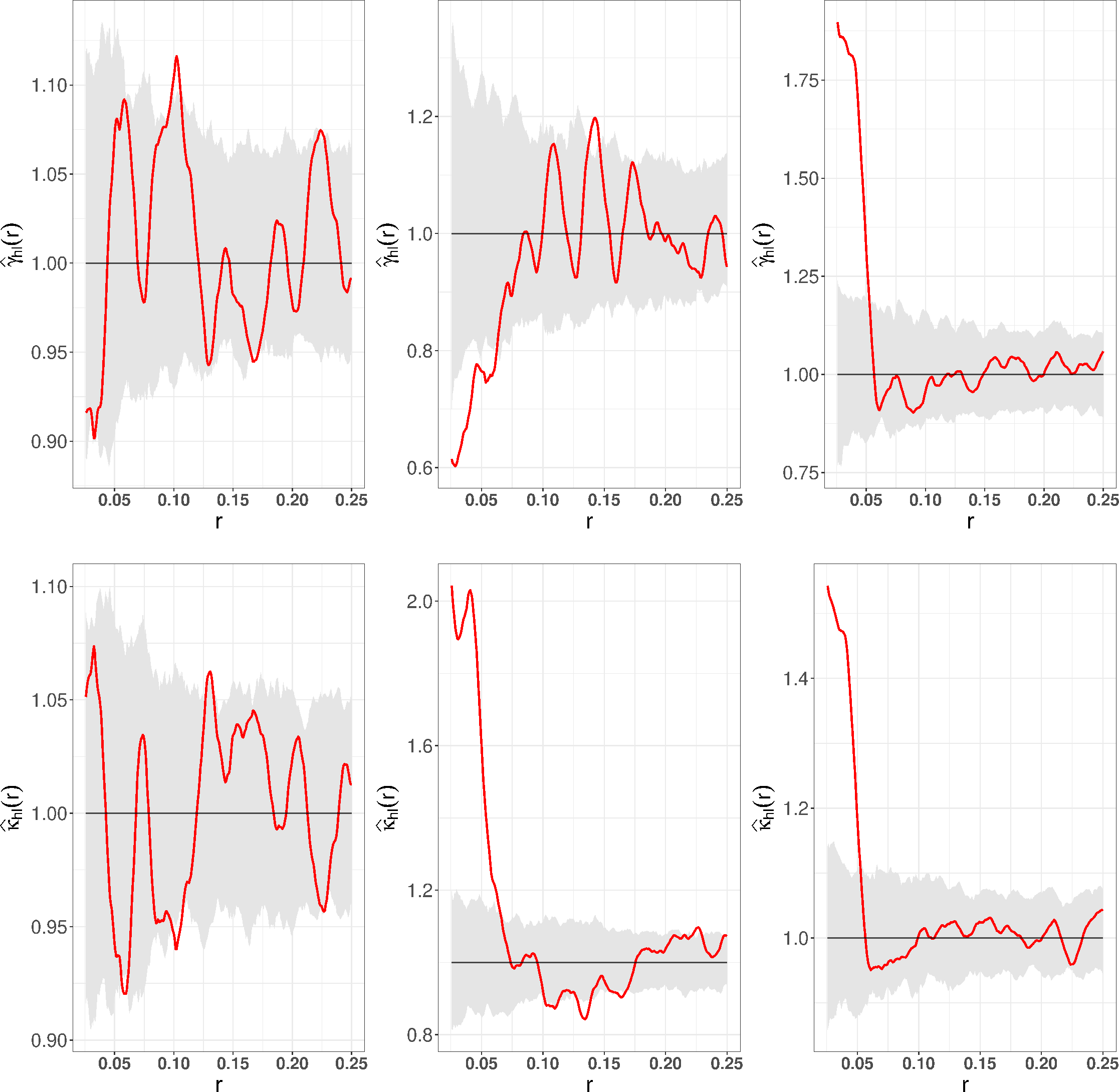}
    \caption{Cross-function mark summary characteristics  for the simulated homogeneous Poisson process on the unit torus with point intensity $\lambda=200$. Cross-function mark variogram (top) and cross-function mark correlation  (bottom)  with no-interaction effects ($J_h(\cdot)=J_l(\cdot)=J(\cdot)$, with c=0) (left), positive inter-function interaction ($J_h(\cdot)=J_l(\cdot)=J(\cdot)$, with $c=0.5$) (central), and negative inter-function correlation $(J_h(\cdot)=J(\cdot)$ and $J_l(\cdot)=0$ with $c=0.5$) (right). Empirical versions of both characteristics are highlighted in red, theoretical values in black. Grey shading shows the fifth-largest and smallest envelope values based on $199$ random simulations according to the null hypothesis of random labeling of functions over fixed point locations.  
    }
    \label{fig:Poisimul}
\end{figure}

\begin{figure}[!h]
    \centering
    \includegraphics[scale=0.3]{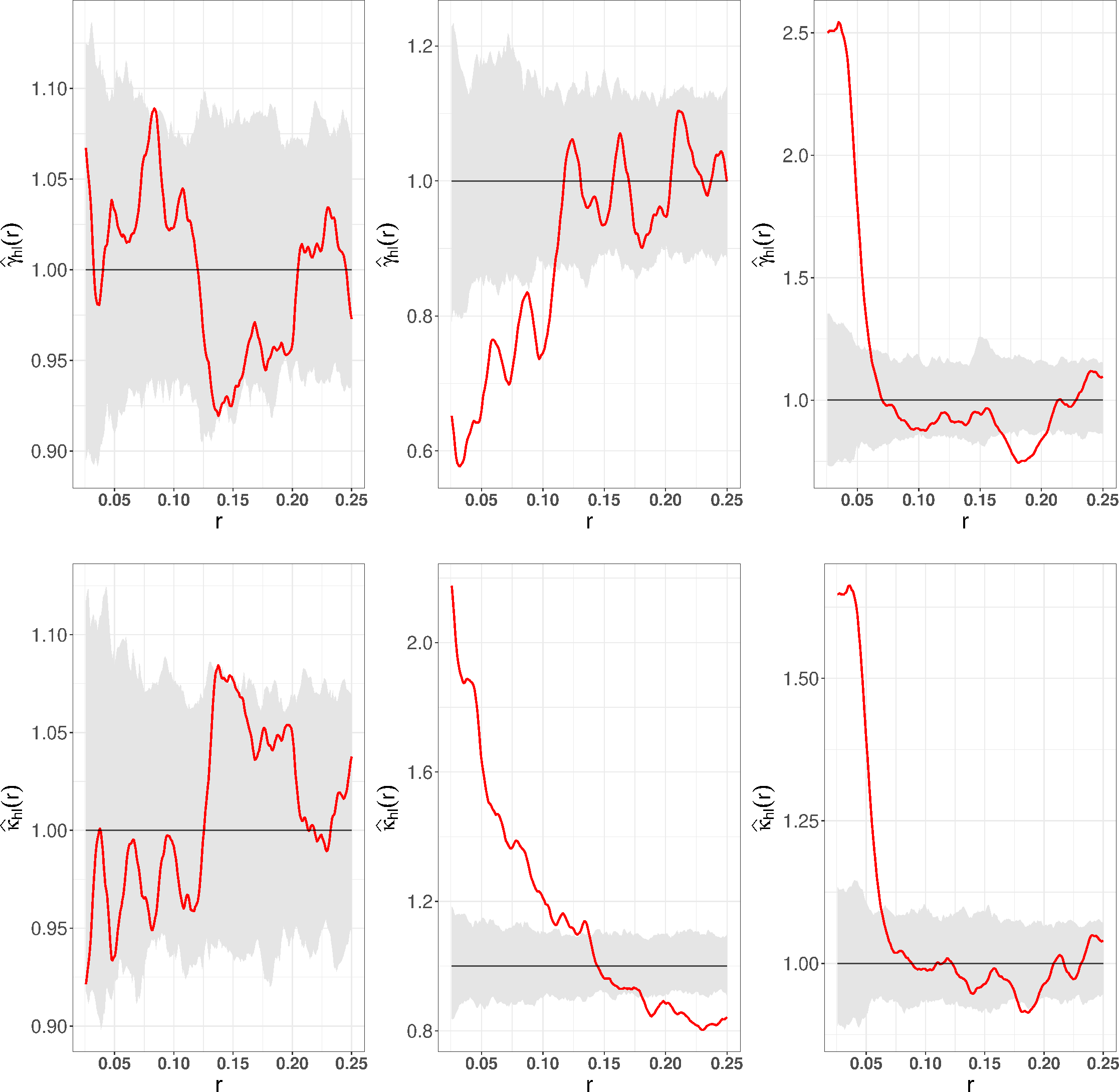}
    \caption{Cross-function mark summary  characteristics for the simulated Thomas process with offspring dispersion parameter $\sigma= 0.04$, parent intensity $\lambda_p = 40$, and $\mu=4$ expected offsprings per parent. Cross-function mark variogram (top) and cross-function mark correlation (bottom)  with no-interaction effects ($J_h(\cdot)=J_l(\cdot)=J(\cdot)$, with c=0) (left), positive inter-function interaction ($J_h(\cdot)=J_l(\cdot)=J(\cdot)$, with $c=0.5$) (central), and negative inter-function correlation $(J_h(\cdot)=J(\cdot)$ and $J_l(\cdot)=0$ with $c=0.5$) (right). Empirical versions of both characteristics are highlighted in red, theoretical values in black. Grey shading shows the fifth-largest and smallest envelope values based on $199$ random simulations according to the null hypothesis of random labeling of functions over fixed point locations.}
    \label{fig:Clusimul}
\end{figure}

Figure \ref{fig:Poisimul} shows the results for the homogeneous Poisson point process scenario with intensity $\lambda=200$. The red lines are the empirical cross-function mark summary characteristics from  a single simulation. The grey shading shows  the fifth-largest and smallest envelope values based on $199$ random simulations according to the null  hypothesis of random labeling of functions over fixed point locations. Here, we consider three correlation function scenarios, namely, spatial independence between functions (left), positive (middle) and negative (right) correlation between functions. This highlights that in absence of interaction between functions, the resulting estimators of both the cross-function mark variogram (\ref{eq:estvario}) (top panels) and the cross-function mark correlation (\ref{eq:estmcor}) (bottom panels) lie within the grey shading area, confirming the spatial independence between functions of type $h$ and $l$. In direct contrast, when assuming spatial positive or negative interaction between curves (central and right panels, respectively), these estimators lie outside this grey shading area, confirming the presence of inter-function dependencies. In particular, under positive correlation of the function-valued marks, the empirical 
cross-function mark variogram lies outside this grey shading area with values smaller than the smallest envelope values, for small $r$ values. This suggests that the positive interacting function-valued marks have less variability than under the independent mark setting. Similarly, under negative correlation between functions, estimators of both the cross-function mark variogram and  the mark correlation lie outside the grey shading area with values  larger than the largest envelope values, for small $r$ values, suggesting negative interactions between functions.

\begin{figure}[!h]
    \centering
    \includegraphics[scale=0.3]{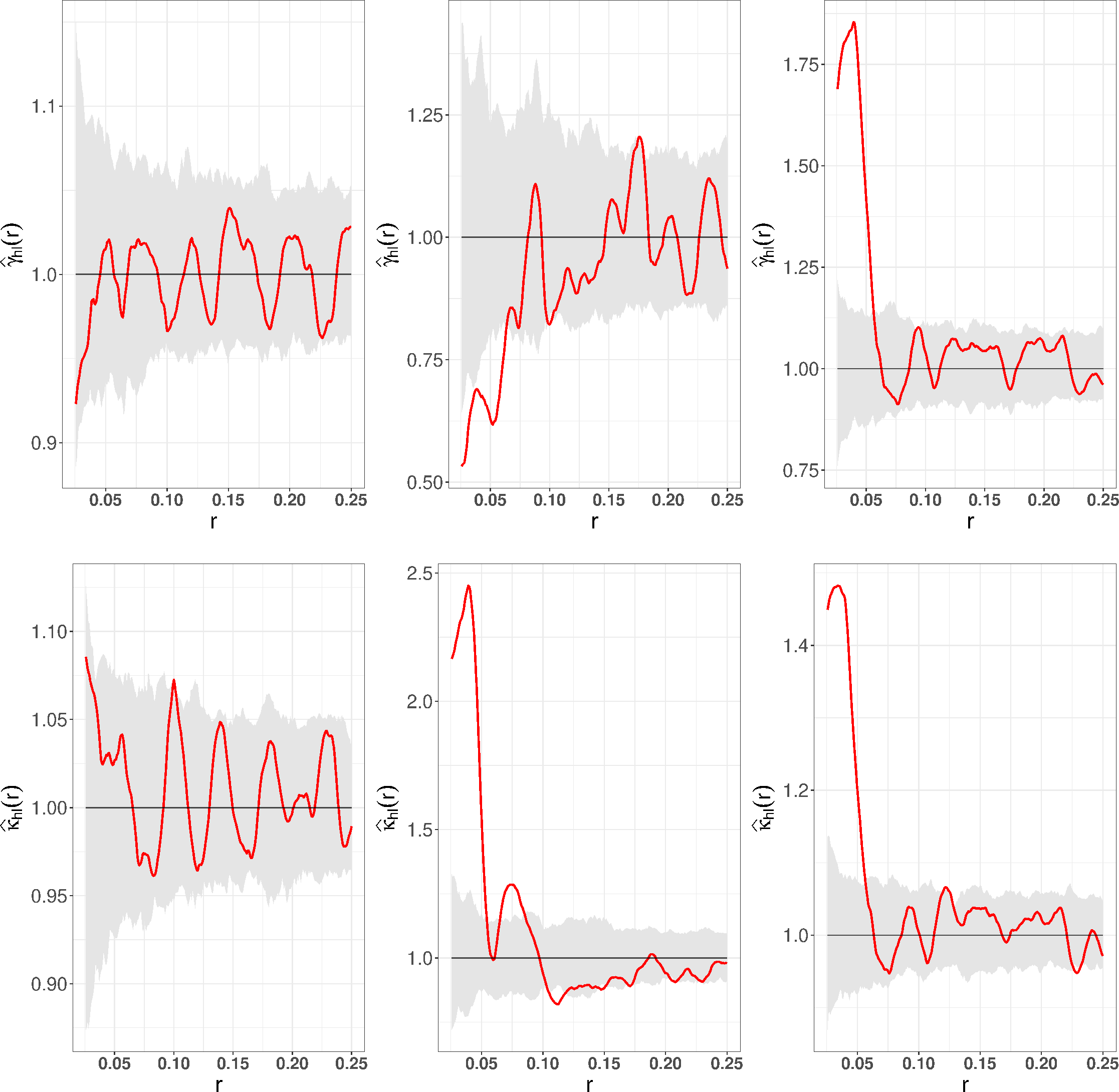}
    \caption{Cross-function mark summary characteristics  for the simulated Strauss process with interaction distance $R_{int}=0.05$ and interaction parameter $q=0.05$. Cross-function mark variogram (top) and cross-function mark correlation (bottom)  with no-interaction effects ($J_h(\cdot)=J_l(\cdot)=J(\cdot)$, with c=0) (left), positive inter-function interaction ($J_h(\cdot)=J_l(\cdot)=J(\cdot)$, with $c=0.5$) (central), and negative inter-function correlation $(J_h(\cdot)=J(\cdot)$ and $J_l(\cdot)=0$ with $c=0.5$) (right). Empirical versions of both characteristics are highlighted in red, theoretical values in black. Grey shading shows the fifth-largest and smallest envelope values based on $199$ random simulations according to the null hypothesis of random labeling of functions over fixed point locations.}
    \label{fig:Strassimul}
\end{figure}

Similar results can be found for the Thomas (Figure \ref{fig:Clusimul}) and the Strauss process scenarios (Figure  \ref{fig:Strassimul}). In absence of inter-function dependencies  both estimators (cross-function mark variogram and mark correlation) lie within the grey shading area, whilst for positive or negative correlation effects between functions these functions lie outside these envelopes. This confirms that the new cross-function mark summary characteristics can detect spatial dependencies between functions of distinct type independently of the spatial structure of the underlying point pattern.

\section{Applications}\label{sec:appl}

\subsection{Application to Swiss tree data}\label{sec:applSwiss}

\begin{figure}[!h]
    \centering
    \includegraphics[scale=0.3]{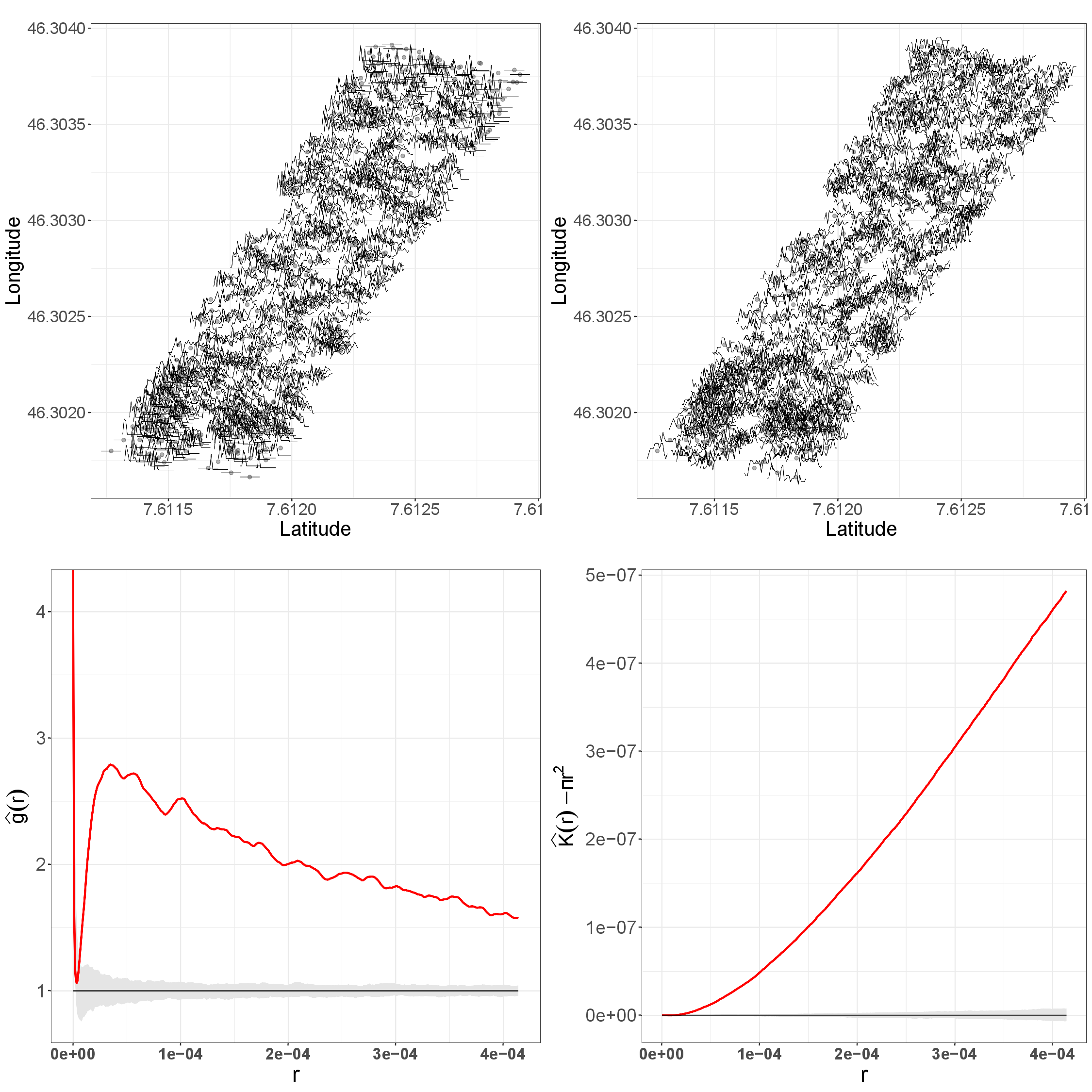}
    \caption{Observed function-valued marks and classic second-order point process summary characteristics of the Swiss tree patterns. Top panel: spatial distribution of Scots pines of the Pfynwald data with observed total crown defoliation (top left) and local pair correlation functions (top right) as function-valued marks. Bottom panel: pair correlation function and theoretical envelopes under the independent mark hypothesis (left), and Ripley's $K(r)$ function minus $r\pi^2$ and theoretical envelopes (right) computed from the point locations. Empirical versions of both characteristics are highlighted in red, theoretical values in black.
 Grey shading shows the fifth-largest and smallest envelope values based on $199$ random simulations according to the null hypothesis of complete spatial randomness (Poisson point randomizations).    
    }
    \label{fig:fctpatternsplotSwiss}
\end{figure}

As a first data application, we consider tree measurements recorded at an annual basis over $14$ years that originates from a long-term irrigation experiment located in \textit{Pfynwald}, the central part of the \textit{Pfyn-Finges} national park in Switzerland \citep{pfynwald:2016}. Initiated in 2003, the experiment aimed to investigate the effect of increased water availability on the individual trees and the ecosystem in a naturally dry Scots pine (Pinus sylvestris L.) forest. The study region  covers an area of 1.2 ha and is located in one of the driest inner-Alpine valleys of the European Alps \citep[see][for detailed summary]{pfynresults}. The data at hand was provided as open data under an Open Database License and has been made available publicly at \url{https://opendata.swiss}. It covers the tree-specific spatial coordinates, the initial assignment into the treatment or control group and different tree characteristics for 900 trees. Form this source, we initially selected the annual total crown defoliation (TCD) from the provided list of tree characteristics and also the exact point locations of the individual tree stands. The TCD parameter is a commonly used parameter in forest monitoring studies to quantify the loss of needles or leaves of a given tree relative to a local reference tree. Within the application, we considered the retrieved TCD information as function-valued tree attribute and assigned it as a mark to the tree locations in a subsequent action. Restricting the data to complete cases, we excluded any trees with incomplete or missing TCD information from the data yielding a final sample of 799 trees with annual TCD records over all 14 years. In a next step, we computed the local pairwise correlation function for all trees of the reduced sample  which describes the contribution of the individual point to the  empirical pair correlation function, i.e. its pair correlation function based local indicator of spatial association. The local information was then used as a second function-valued mark in our application such that each tree was marked by two distinct function-valued quantities. The resulting point pattern with both function-valued marks and classic second-order summary characteristics of the points are shown in Figure \ref{fig:fctpatternsplotSwiss}. While not considered here, we note that the data also allows for cross-function cross-type versions as outlined on Section \ref{sec:multipoints} by taking additionally the tree-specific assignment into treatment or control group into account. Such advanced mark characteristics might help to investigate the complex interplay of the TCD and local pair correlation function curves with the effect of additional water supply.

As expected by the large number of trees, the sampled point pattern reflects some clear structure and a tendency of clustering among the points. This impression is supported by the pair correlation function (left bottom panel) and also Ripley's $K$ function (right bottom panel) which show a clear positive shift of the empirical curves from the theoretical lines under the complete spatial randomness hypothesis which indicates a clear tendency of clustering.

\begin{figure}[!h]
    \centering
    \includegraphics[scale=0.3]{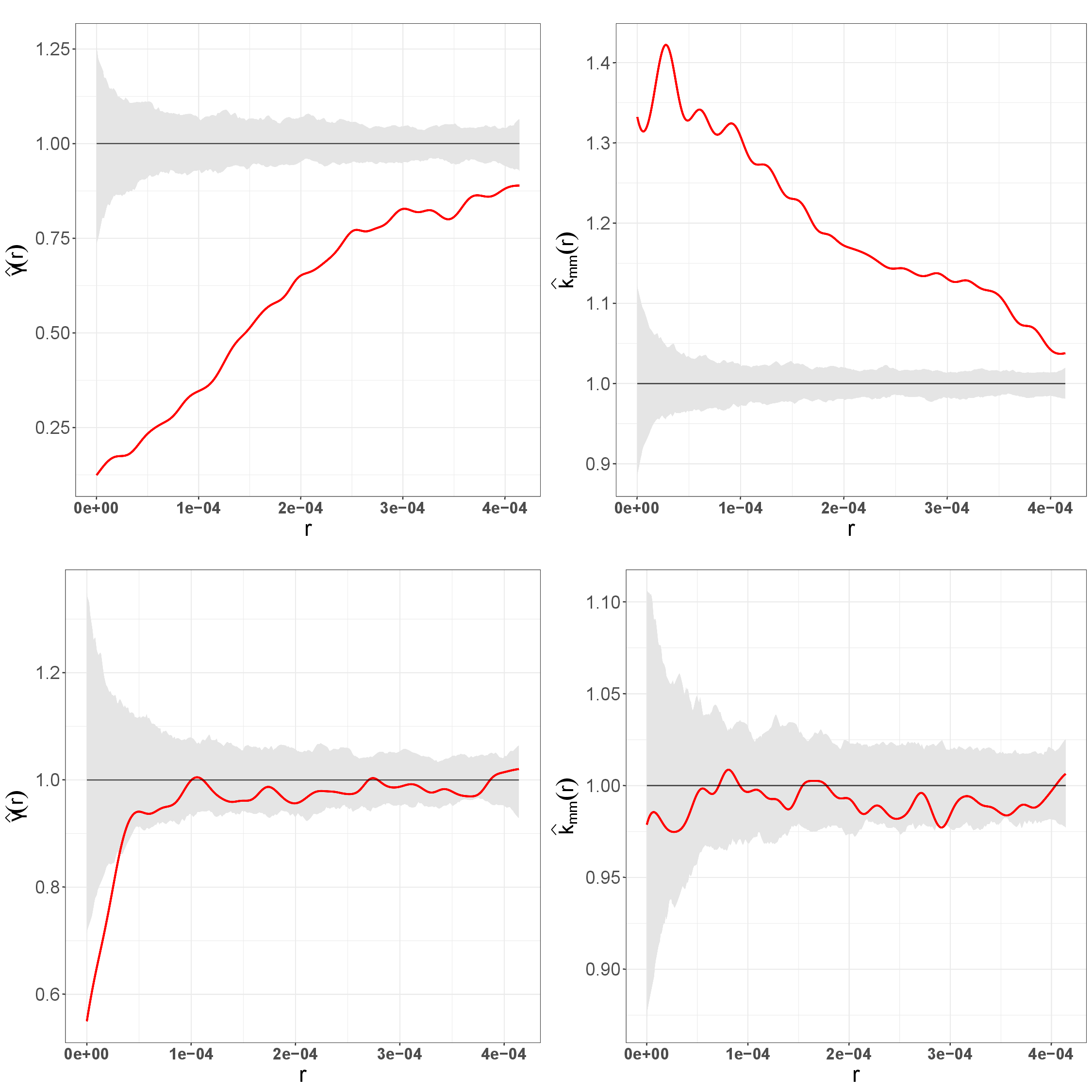}
    \caption{Classic mark summary characteristics for the Pfynwald tree data with averaged function-valued point attributes treated as scalar-valued marks. Mark variogram and mark correlation functions for the mean TCD (top) and mean local pair correlation function (bottom).  Empirical versions of both characteristics are highlighted in red, theoretical values in black.
  Grey shading shows the fifth-largest and smallest envelope values based on $199$ random simulations according to the null hypothesis of random labeling of marks over fixed point locations.  
    }
    \label{fig:MarkCharsSwiss}
\end{figure}

Next, to evaluate the findings of the proposed auto- and cross-function summary characteristics with the classic summary characteristics for scalar-valued marks commonly used at present, we transformed the function-valued marks into function-wise averages and computed the mark variogram and Stoyan's mark correlation function from the averaged quantities (see Figure \ref{fig:MarkCharsSwiss}).  The empirical versions of both mark characteristics  show a clear deviation from the theoretical envelopes for the average TCD (top panel). While the mark variogram (left top panel) suggests that the mean TCD values exhibit less pairwise variation as expected under the independent mark hypothesis,  we found a clear positive shift of the empirical pairwise product of TCD averages as considered by the mark correlation function  (right top panel) from the theoretical envelopes. In comparison with the TCD, both empirical mark characteristics show almost no deviations from the independent mark hypothesis in case of the averaged local pair correlation function (bottom panels). Except only some negative shift of the mark variogram (left) at small distances, both estimated characteristics are covered by the envelopes.     

\begin{figure}[!h]
    \centering
    \includegraphics[scale=0.3]{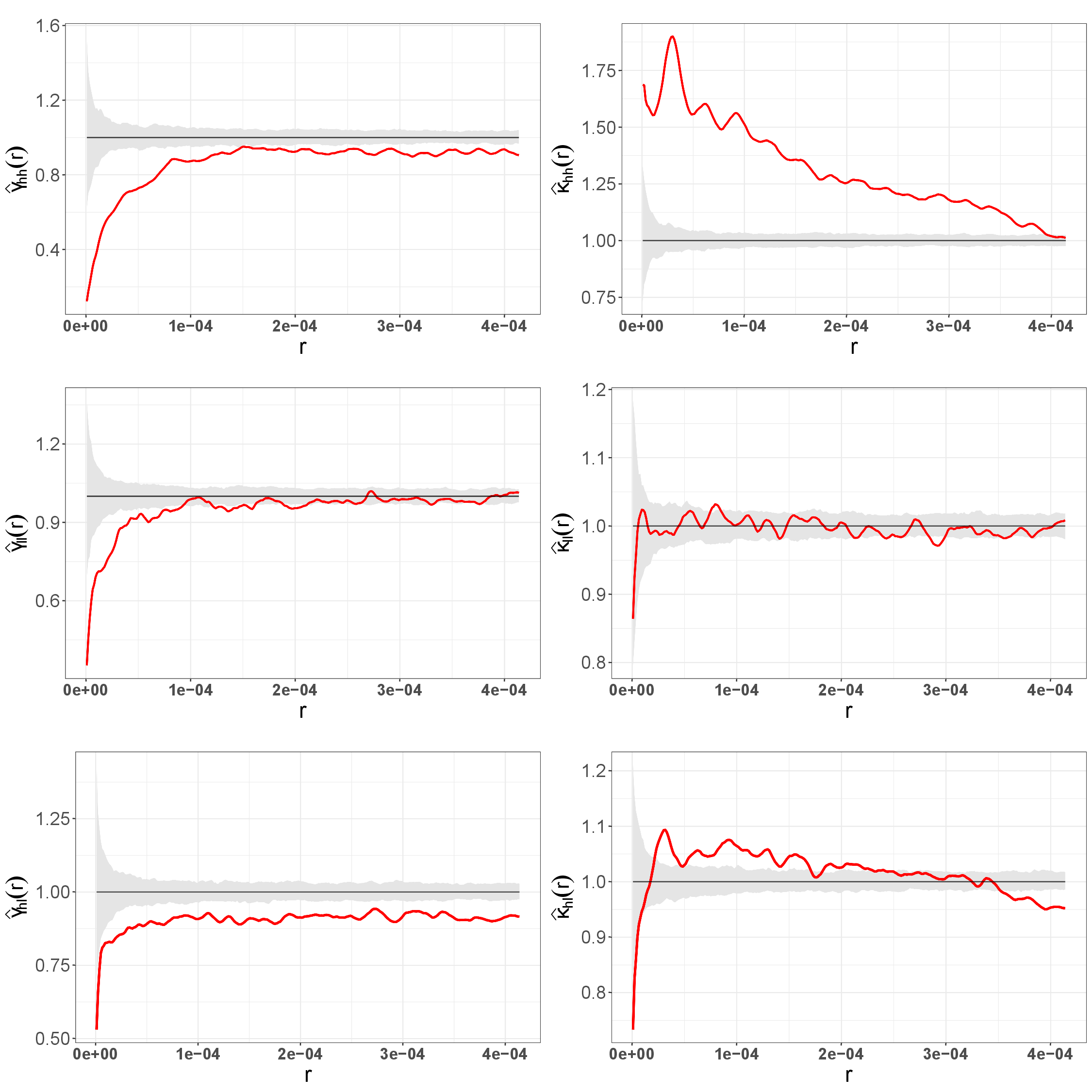}
    \caption{Auto- and cross-function mark summary characteristics computed from the Swiss tree data. Top: auto-function mark variogram (left) and mark correlation function (right) of the  total crown defoliation  curves. Central panel: auto-function mark variogram (left) and mark correlation function (right) of the local pair correlation functions. Bottom: cross-function (left) and mark correlation function (right) between the total crown defoliation and the local pair correlation functions. Empirical versions of both characteristics are highlighted in red, theoretical values in black. 
Grey shading shows the fifth-largest and smallest envelope values based on $199$ random simulations according to the null hypothesis of random labeling of functions over fixed point locations. }
    \label{fig:fctCharsSwiss}
\end{figure}

Different from the classic mark characteristics, all auto- and cross-function mark variograms and correlation functions of Figure \ref{fig:fctCharsSwiss}, except the auto-function mark correlation of the local pair correlation (central right panel), show significant results. As already indicated by the classic characteristics, the top panel corresponding to the TCD curves reflects again a negative deviation of the empirical auto-function mark variogram  (left top panel) contrasted with a clear  positive shift of the empirical auto-function mark correlation function (right top panel) from the theoretical lines under the independent mark hypothesis. This indicates that the observed  TCD curves show less spatial variation among pairs of neighbouring points. At the same time, the product of the TCD curves clearly exceeds the expected case, i.e. the non-spatial functional mean squared. For the central panels showing the auto-function characteristics computed from the local pair correlation functions, the auto-function mark variogram (left) again  suggests smaller variation between the function-valued marks compared to  the independent mark setting for some small distances. Finally, looking at the cross-function characteristics of the TCD and local pair correlation curves, both results show a clear variation from the independent mark envelopes.  This would imply that the pairwise spatial variation of both functions is smaller than under the limiting case where the cross-function variogram is equal to the covariance, whereas the pairwise product of the two marks exceeds the limiting case in which the pairwise product of the two marks approaches the product of the functional means $\mu_h$ and $\mu_l$.  

\subsection{Application to Spanish labour data}

\begin{figure}[!h]
    \centering
    \includegraphics[scale=0.3]{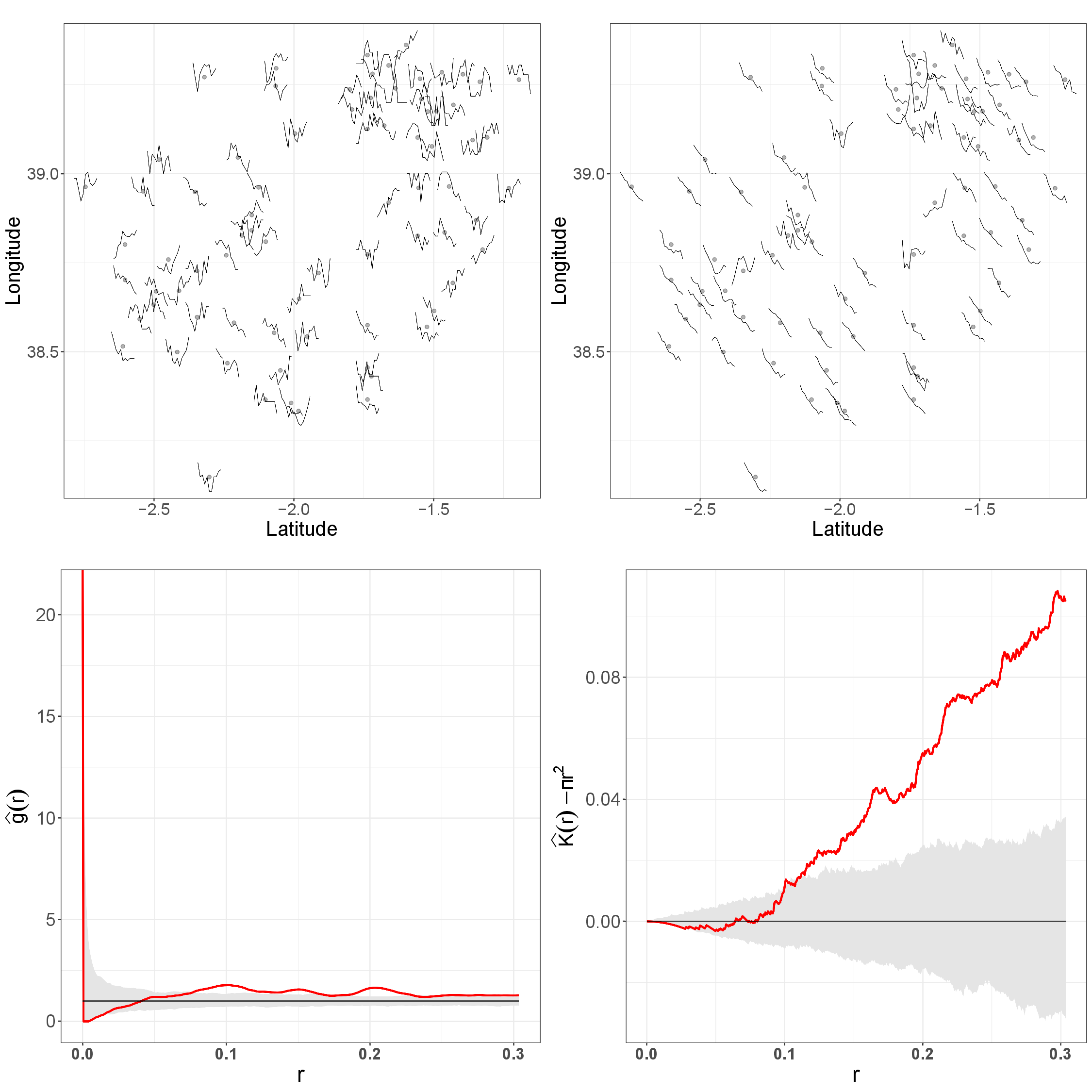}
    \caption{Observed function-valued marks and classic second-order point process summary characteristics computed from 84 municipalities of the province of Albacete. Top panel: spatial distribution of Spanish municipalities and the yearly differences to the reference year 2022 for the business (top left) and population (top right) records as function-valued marks. Bottom panel: pair correlation function and theoretical envelopes under the independent mark hypothesis (left), and Ripley's $K(r)$ function minus $r\pi^2$ and theoretical envelopes (right) computed from the point locations. Empirical versions of both characteristics are highlighted in red, theoretical values in black.
    Grey shading shows the fifth-largest and smallest envelope values based on $199$ random simulations according to the null hypothesis of complete spatial randomness (Poisson point randomizations). }   
    \label{fig:fctpatternsplot}
\end{figure}

As second example of a spatial point pattern with bivariate function-valued marks we considered data on the total number of companies as of January, 1st and the number of residents recorded annually at municipality level for the period from 2012 to 2022. The data originated from the official data reports released by the National Statistics Institute of Spain (INE) and was made publicly available at \url{www.ine.es}. The business information was derived from the official business register of INE and corresponds to the total  number of local  companies over different economic sectors. The local assignment of the companies to exactly one municipality was performed by INE in a pre-processing step using the registered business address information to avoid potential inconsistencies in case of regionally wide spreading business locations, e.g. factories or business facilities of one company in several distinct municipalities. From the provided data we initially selected  a sample of 87 municipalities that fall into the boundaries of \textit{Albacete}, a Spanish province on \textit{La Mancha} (the Spanish Plateau). The area of \textit{La Mancha} is located southeast of Madrid and is characterised by a homogeneous climate and population density, and as such, has been treated as a particular instance of a homogeneous spatial point process in the literature  \citep[see e.g.][for some applications]{GlasTobler:71, ripley77, ripley88,Chiu2013}. From the selected file we excluded the municipalities of \textit{Masegoso}, \textit{Montalvos} and \textit{Villa de Ves} in a subsequent action for which no economic information was available. This yielded a final sample of 84 Spanish municipalities to which we applied the following pre-processing. In a first step, we derived the exact spatial location of the centroids for each  municipality in the sample and assigned the corresponding pair of coordinates to the data. Next, we generated two function-valued attributes from the provided local business and population statistics by computing the pointwise yearly differences between the values from 2012 to 2021 and the reference records of 2022. As such, both generated marks express the annual change in the size (resp. number) of the local business sector (resp. population) with respect to 2022.  All information was then transformed into a spatial point process with function-valued marks in a final step. The generated point pattern and classic second-order point process summary characteristics of the points are shown in Figure \ref{fig:fctpatternsplot}. 

Different from the Swiss tree data example, the Spanish point pattern appears to be less dense and clustered with both   constructed business (\ref{fig:fctpatternsplot}, upper left panel) and population (\ref{fig:fctpatternsplot}, upper right panel) showing some heterogeneity. Both the empirical pair correlation function (bottom left panel) and Ripley's $K$ function minus $r\pi^2$ indicate a clear tendency to clustering for the point locations, which supports the visual impression.

\begin{figure}[!h]
    \centering
    \includegraphics[scale=0.3]{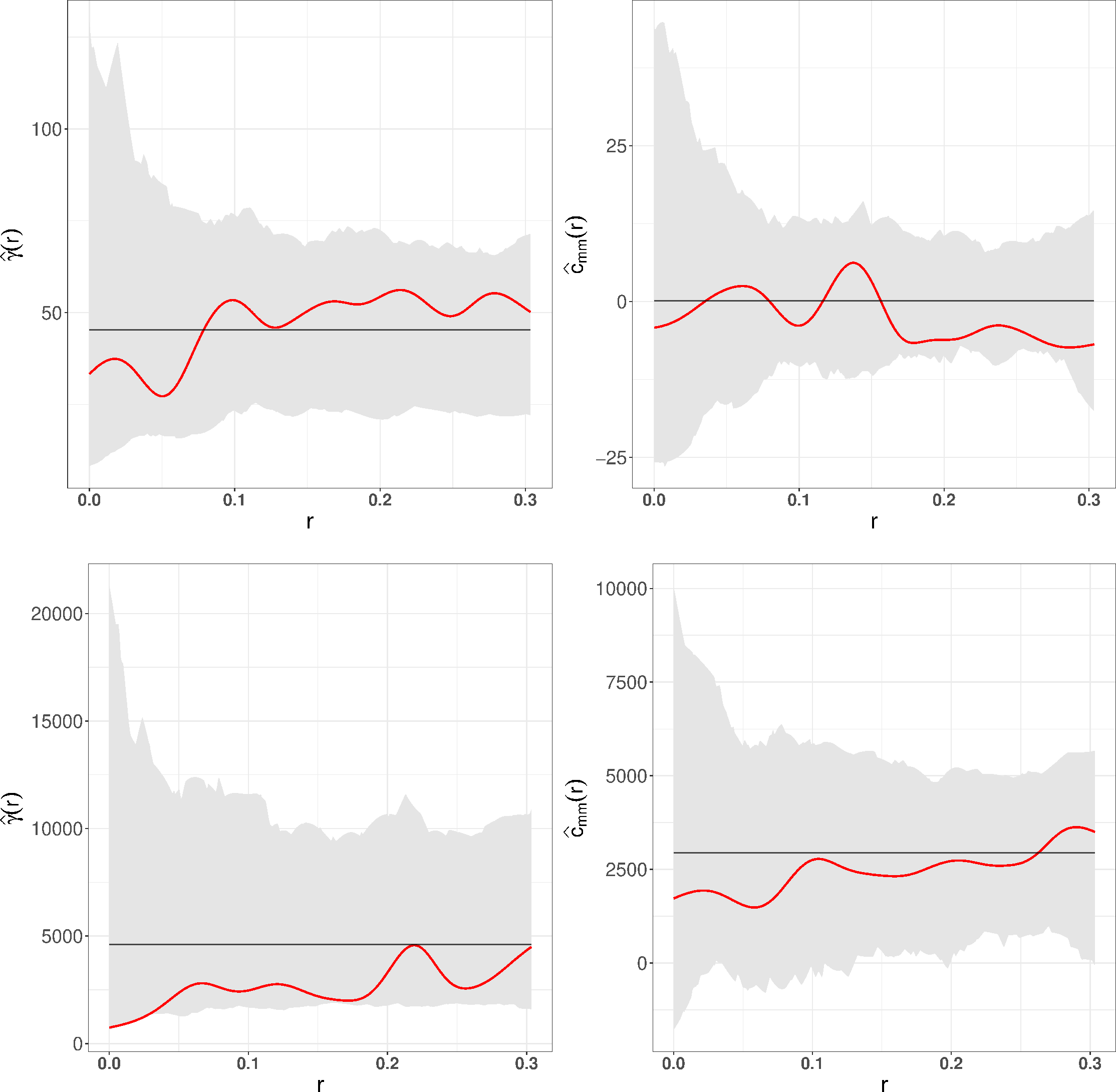}
    \caption{Classic mark summary characteristics for the Spanish municipality data computed from the averaged business and population information with averaged function-valued point attributes treated as scalar-valued marks. Mark variogram and conditional mean product of marks for the mean business (top) and mean population function (bottom).  Empirical versions of both characteristics are highlighted in red, theoretical values in black.
     Grey shading shows the fifth-largest and smallest envelope values based on $199$ random simulations according to the null hypothesis of random labeling of marks over fixed point locations.}
    \label{fig:INEScalarChr}
\end{figure}

As for Section \ref{sec:applSwiss}, we computed the means from the business and population curves and used the scalar information as input for classic mark summary characteristics (see Figure \ref{fig:INEScalarChr}). Due to the presence of negative mark values, the unnormalised version, i.e. the conditional mean product of marks $c_{mm}(r)$, was computed instead of the mark correlation function $k_{mm}(r)$. 
For the averaged business  variation (top panels) and the mean variation of the population (bottom panels) both the mark variogram and mark correlation function are completely covered by the envelopes, supporting the independent mark hypothesis.    

\begin{figure}[!h]
    \centering
    \includegraphics[scale=0.35]{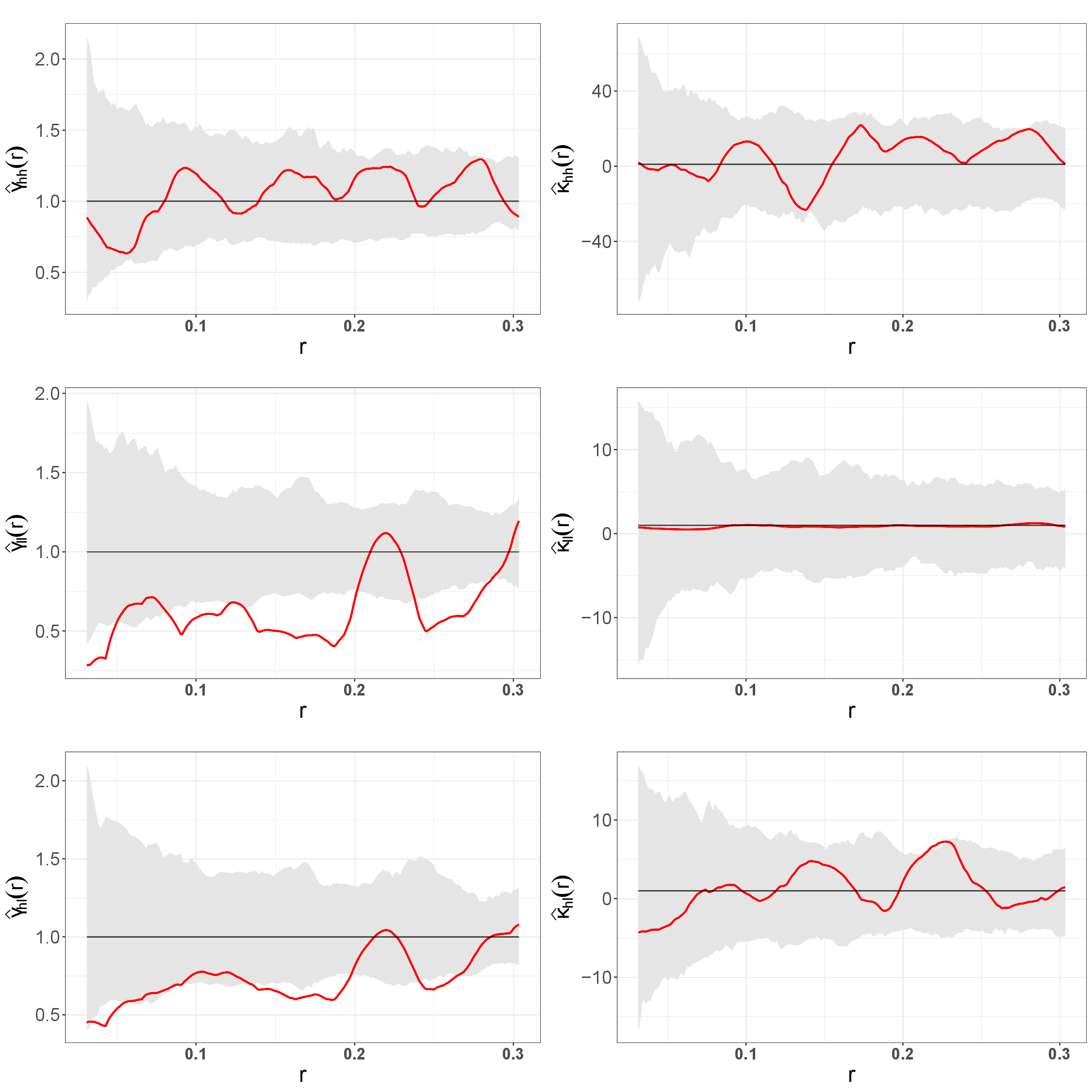}
    \caption{Auto- and cross-function mark summary characteristics computed from the Spanish municipality data. Top: auto-function mark variogram (left) and mark correlation function (right) of the business curves. Central panel: auto-function mark variogram (left) and mark correlation function (right) of the population curves. Bottom: cross-function (left) and mark correlation function (right) between the business and population curves. Empirical versions of both characteristics are highlighted in red, theoretical values in black. Grey shading shows the fifth-largest and smallest envelope values based on $199$ random simulations according to the null hypothesis of random labeling of functions over fixed point locations. 
    }
    \label{fig:fctChrINE}
\end{figure}

Comparing these these findings with the results of the auto- and cross-function mark characteristics depicted in Figure \ref{fig:fctChrINE}, the independence hypothesis is not supported by the auto-function mark variogram for population (central left panel) and the cross-function mark variogram of business and population (bottom left). Both functions would instead suggest less pairwise variation between the generated populations differences resp. business and population differences as expected under the independent mark assumption. As both mark summary characteristics for the change in size of the business sectors (top panels) are included in the envelopes, the significant results of the cross-function mark variogram seem to be driven mostly be the pairwise variation of the population curves.         

\section{Conclusions}\label{sec:final}

This paper proposes an immense variety of different mark summary characteristics which allow to decide on potential structure of the marks within highly challenging spatial point process scenarios. Including cross-function, multi-function and corresponding mark weighted versions of well-established  mark summary characteristics, the extended methods are providing a suitable statistical toolbox for the analysis of spatially aligned function-valued quantities for a plethora of potential applications. Formalised through generalisations of classical test functions to the (multivariate) function-valued mark scenario, the proposed characteristics are well embedded into the statistical literature and methodology for spatial point patterns with real-valued marks and allow for similar interpretations.     

The considered estimators are natural extensions to the complex function-valued cases of those proposed in  \citet{comas2008METMA,Comas2011, Comas2013} and are technically supported by the theoretical treatments in \citet{Ghorbani2020}. We rely on this latter contribution to support the behaviour of the proposed estimators. However, there are a number of doors open in both, theoretical and inferential aspects, that starting from our developments can go further in providing tools for complex mark structures. One such example we can think of is the case when we have trajectories restricted to a network-based topology. Here more topological arguments are needed to be considered when developing further tools and their estimators.

\section*{Acknowledgements}

The authors gratefully acknowledge financial support through the German Research Association and the
Spanish Ministry of Science. Matthias Eckardt was funded by the Walter Benjamin grant 467634837 from the German Research Foundation. Jorge Mateu was partially funded by grant PID2019-107392RB-I00, from the Spanish Ministry of Science. Carles Comas was partially funded by grant PID2020-115442RB-I00 from MCIN/AEI/ 10.13039/501100011033 (Spanish Ministry of Science)

\bibliographystyle{ecta}
\bibliography{fmspp}
\end{document}